\documentclass[12pt]{article}
\usepackage[latin9]{inputenc}
\usepackage{float}
\usepackage{amsmath}
\usepackage{amssymb}
\usepackage{graphicx}
\usepackage{geometry}
\usepackage[numbers]{natbib}
\usepackage[bookmarks=false,
 breaklinks=false,pdfborder={0 0 1},backref=section,colorlinks=false]
 {hyperref}

\makeatletter

\providecommand{\tabularnewline}{\\}

\@ifundefined{date}{}{\date{}}
\usepackage{authblk}
\author{Jos\'e Weberszpil\thanks{Email: josewebe@gmail.com}}
\affil{Universidade Federal Rural do Rio de Janeiro, UFRRJ-DEFIS/ICE\\
BR-465, Km 7 Serop\'edica-Rio de Janeiro
CEP: 23.897-000}
\author {Ralf Metzler\thanks{Email:ralf.metzler@uni-potsdam.de}}
\affil{Institute of Physics and Astronomy, University of Potsdam,\\ Karl-Liebknecht-Str. 24/25, 14476 Potsdam-Golm, Germany}

\date{\today}

\makeatother

\begin{document}
\title{An Extended Model of Non-Integer-Dimensional Space for Anisotropic
Solids with $q-$Deformed Derivatives}
\maketitle
\begin{abstract}
We propose a non-integer-dimensional spatial model for anisotropic
solids by incorporating a $q-$deformed derivative operator, inspired
by the Tsallis nonadditive entropy framework. This generalization
provides an analytical framework to explore anisotropic thermal properties,
within a unified and flexible mathematical formalism. We derive explicit
expressions for the phonon density of states and specific heat capacity,
highlighting the impact of the deformation parameter $q$ on the thermodynamic
behavior. We apply the model to various solid-state materials, achieving
excellent agreement with experimental data across a wide temperature
range, and demonstrating its effectiveness in capturing anisotropic
and subextensive effects in real systems. Beyond providing accurate
fits, we anchor the $q-$deformation in a microscopic disorder/kinetics
exponent $\mu$ emerging from conformable dynamics, thereby linking
nonextensive statistics to measurable heterogeneity and memory effects.
\end{abstract}

\section{Introduction}

Anisotropic materials exhibit direction-dependent physical properties
that fundamentally challenge conventional thermodynamic models based
on three-dimensional isotropic model approaches \citep{Ziman1960,Ashcroft1976,Kittel2005}.
These materials, ranging from layered van der Waals crystals \citep{Novoselov2016,Geim2013}
to nanostructured composites \citep{Ajayan2003,Coleman2011}, demonstrate
thermal transport phenomena that cannot be adequately described by
classical Debye theory alone \citep{Debye1912,Ashcroft1976}. The
directional asymmetries in such systems often arise from underlying
microstructural heterogeneities, crystalline anisotropy, or topological
constraints that effectively reduce the dimensionality of phonon propagation
\citep{nakayama1994dynamical,Balandin2011,Alexander1982,Orbach1986}.

In 1990, He proposed a groundbreaking framework to address this challenge
by mapping anisotropic systems to effectively isotropic entities embedded
in a non-integer-dimensional space characterized by a Hausdorff dimension
$\alpha$~\citep{he1990}. This geometric approach provides a mathematically
elegant method to encode anisotropic effects through dimensional reduction,
enabling the application of conventional statistical mechanics to
complex materials with modified density of states (DOS).

However, many anisotropic solids often exhibit additional complexities
beyond simple dimensional reduction. Many materials display nonlinear
thermal responses, memory effects, and correlations that suggest deviations
from extensive thermodynamic behavior \citep{tsallis1988,Tsallis2009,Tsallis2019}.
These phenomena are particularly pronounced in nanostructured materials
\citep{pradhan2008specific}, where surface effects, quantum confinement,
and structural disorder introduce non-trivial statistical signatures
\citep{MetzlerKlafter2000,MetzlerKlafter2004}.

Recent advances in our microscopic understanding of $q-$deformed
derivative operators \citep{weberszpil2015connection,weberszpil2021entropy,liang2019fractal,weberszpil2016variational,weberszpil2017generalized,rosa2018dual,sotolongo2023fractal,xu2017spatial}
and other deformed derivatives, such as conformable derivatives \citep{weberszpil2025microscopic}
have demonstrated their effectiveness in modeling systems with anomalous
diffusion, non-Gaussian fluctuations, and nonextensive behavior. Building
on this theoretical foundation, the present work extends He's non-integer-dimensional
framework by incorporating $q-$deformed derivatives inspired by Tsallis'
nonadditive entropy formalism \citep{tsallis1988}.

The $q-$deformed derivative operator, was defined as 
\begin{equation}
D_{q}[f](x)=[1+(1-q)x]\frac{df}{dx},\label{eq:q-Deformed-Derivative}
\end{equation}
by Borges in the context of nonextensive thermostatistics \citep{Borges2004}.
In the original construction, $x$ is dimensionless. In the present
context, $x$ denotes a suitably scaled frequency or energy variable,
such as $\omega/\omega_{0}$ or $\beta E$, so that the factor $[1+(1-q)x]$
remains dimensionless and does not affect the units of the resulting
physical observable.

This deformed derivative offers a systematic way to incorporate nonlinear
corrections that effectively encode nonlocal interactions and memory
effects, as typically encountered in complex materials. In the limit
$q\to1$, the operator reduces to the ordinary derivative, guaranteeing
consistency with standard thermodynamic descriptions. Moreover, the
$q$-deformed operator $D_{q}$ furnishes a convenient framework for
modeling systems in which anisotropy depends explicitly on position
or energy. Examples include quantum wells and superlattices with engineered
anisotropy \citep{Hyldgaard1997,Simkin2000}, fractals and hierarchically
structured materials such as silica aerogels and porous silicon \citep{Schaefer1986,Perez2000},
nanocomposites with interface-dominated transport \citep{pradhan2008specific};
and materials exhibiting non-Debye phonon behavior \citep{Alexander1982}.

The combination of non-integer-dimensional space embedding with $q-$deformation
creates a unified analytical framework capable of describing both
geometric anisotropy and statistical nonextensivity within a single
mathematical formalism. As will be shown in the present work, this
approach enables closed-form expressions for modified DOS and specific
heat capacity that can be directly validated against experimental
data.

Classical Debye and He's fractional-dimensional approaches capture
much of the low temperature behavior, but anomalous transport and
memory effects in complex solids require the use of fractional/CTRW
tools and fractional Fokker-Planck dynamics. For a comprehensive background
on anomalous diffusion and fractional dynamics we refer to \citep{MetzlerKlafter2004,MetzlerKlafter1999PRL}.
The memory-induced anomalous response discussed here is consistent
with the CTRW/fractional kinetics and FFPE paradigm.

\section{Deformed Thermodynamic Model in Non-Integer-Dimensional Space}

Following He's original approach~\citep{he1990}, an anisotropic
medium can be mapped onto an isotropic system embedded in a non-integer-dimensional
space with Hausdorff dimension $\alpha$.

The $q-$deformed derivative operator employed here is closely related
to those previously discussed in the context of anomalous diffusion
and generalized entropy production \citep{weberszpil2015connection,weberszpil2021entropy,weberszpil2016variational,weberszpil2017generalized}.
These earlier formulations demonstrated how such operators encode
effective thermodynamic deformation and memory effects in a variety
of physical settings. 

\subsection{Non-integer-dimensional embedding with deformed derivatives}

We reinterpret the momentum-space measure and excitation statistics
in terms of $q$-deformed calculus \citep{weberszpil2015connection,Borges2004}.
In the Hausdorff-conformal approach, the derivative operator acting
on a function $f$ may be written as

\begin{equation}
D_{\alpha}[f](k)=\Omega_{\alpha}k^{\alpha-1}\frac{df}{dk},\qquad\Omega_{\alpha}=\frac{2\pi^{\alpha/2}}{\Gamma(\alpha/2)},
\end{equation}
where $k=|\mathbf{k}|$ is the magnitude of the wave vector in momentum
space. The factor $k^{\alpha-1}$ is the radial weight induced by
the hyperspherical surface measure in $\alpha$-dimensional space
\citep{he1990,stillinger1977,Chen2006}. This construction is consistent
with the Hausdorff-measure formalism for non-integer-dimensional spaces
and with the axiomatic framework of non-integer-dimensional integration
\citep{he1990,stillinger1977,Falconer2003,Hausdorff1918}. We note
that foundational aspects of non-integer-dimensional space theory
can be traced back to seminal work by Stillinger \citep{stillinger1977},
which established the mathematical framework for non-integer-dimensional
calculus in physical systems.

In the condensed-matter setting, this type of non-integer-dimensional
description is also consistent with fractional-dimensional models
for anisotropic solids.

\paragraph{Radial meaning of $k$ and origin of the factor $k^{\alpha-1}$.}

Throughout this work, $k:=|\mathbf{k}|\ge0$ denotes the \emph{radial}
magnitude in an $\alpha$-dimensional (generally non-integer) isotropic
momentum space. Accordingly, the Hausdorff measure in hyperspherical
coordinates reads 
\begin{equation}
d^{\alpha}k\;=\;S_{\alpha-1}\,k^{\alpha-1}\,dk,\qquad S_{\alpha-1}=\frac{2\pi^{\alpha/2}}{\Gamma(\alpha/2)},\label{eq:hausdorff_measure_radial}
\end{equation}
where $S_{\alpha-1}$ is the surface area of the unit $(\alpha-1)$-sphere,
so that the factor $k^{\alpha-1}$ is the usual radial Jacobian in
$\alpha$ dimensions. Because $k$ is by definition nonnegative (radial
coordinate), no absolute value $|k|^{\alpha-1}$ is required in \eqref{eq:hausdorff_measure_radial}.
If one were to write the measure in Cartesian components over $\mathbb{R}^{\alpha}$,
the integrand is rotation invariant and the reduction to radial integration
yields exactly the factor $k^{\alpha-1}$.

In this sense, the factor $k^{\alpha-1}$ is \emph{not} an ad hoc
ingredient but the Jacobian associated with isotropic mode counting
in Hausdorff dimension $\alpha$, consistent with the fraction-dimensional
construction used by \citep{he1990} and the general non-integer-dimensional
calculus framework \citep{stillinger1977}. %================== END PATCH M1 ==================

\subsection{Physical interpretation of the $q-$deformed framework}

The introduction of $q-$deformed derivatives in the context of anisotropic
solids requires careful physical interpretation to justify its application
beyond mere mathematical formalism. The deformation parameter $q$
encodes deviations from extensive statistical behavior that arise
naturally in systems with:

\textbf{Geometric constraints:} In materials with fractal or hierarchical
microstructures, the effective phase space accessible to excitations
becomes restricted. The $q$-deformation captures this reduced accessibility
through a position-dependent weighting of the derivative operator,
effectively encoding the geometric constraints imposed by the material's
internal structure \citep{he1990,Falconer2003}.

\textbf{Correlation effects:} Anisotropic materials often exhibit
long-range correlations due to directional bonding, layered structures,
or coherent interfaces. The factor $[1+(1-q)x]$ in Eq. (\ref{eq:q-Deformed-Derivative})
introduces memory-like effects where the local response depends on
the energy scale, mimicking the influence of correlations on thermodynamic
quantities \citep{tsallis1988,weberszpil2015connection}.

\textbf{Interface and surface contributions:} In nanostructured anisotropic
materials, surface and interface effects become significant. The $q$-deformation
naturally accounts for these contributions by modifying the DOS in
an energy-dependent manner, reflecting the altered vibrational spectrum
near interfaces \citep{pradhan2008specific,maynard1985}.

\textbf{Non-Debye phonon behavior:} Many anisotropic materials exhibit
phonon dispersion relations that deviate from the linear Debye approximation.
The $q-$deformed DOS captures these deviations through its nonlinear
energy dependence, providing a phenomenological description of complex
phonon interactions.

The mathematical structure of the $q-$deformed framework ensures
that these physical effects are encoded in a thermodynamically consistent
manner. The approach preserves fundamental thermodynamic relationships
while allowing for systematic deviations from extensive behavior,
making it particularly suitable for modeling real materials with complex
internal structures. In Ref. \citep{weberszpil2025microscopic}, the
reader can find more detailed explanation on microscopic origins of
deformed derivative and its connection \citep{weberszpil2015connection}.

Applications of Tsallis nonextensive statistical mechanics have demonstrated
remarkable success in describing systems with long-range correlations,
multifractal structures, and memory effects \citep{Tsallis2009,Tsallis2019,Naudts2011,Abe2001}.
The framework has been particularly effective in condensed matter
systems exhibiting power-law distributions, anomalous diffusion, and
non-Gaussian fluctuations \citep{Boon2005,Pickup2009}. In this context
we note that the use of $q$-deformed calculus (including $q$-derivatives)
and the broader nonadditive-entropy framework has been developed \citep{Borges2004}
and applied extensively beyond the original Tsallis proposal \citep{tsallis1988}.

\subsection{Derivation of the cumulative number of states in non-integer-dimensional
space}

To determine the cumulative number of quantum states $G(E)$ up to
a given energy $E$ in a non-integer-dimensional space of Hausdorff
dimension $\alpha$, we proceed as follows.

The number of quantum states with wavevector magnitude less than $k$
is given by \citep{stillinger1977,Kittel2005,Pathria2011,he1990}
\begin{equation}
G(E)=\frac{1}{(2\pi)^{\alpha}}\int_{|\vec{k}|\leq k(E)}d^{\alpha}k,
\end{equation}
where $d^{\alpha}k$ denotes the infinitesimal volume element in momentum
space with Hausdorff dimension $\alpha$.

The volume of a hypersphere of radius $k$ in $\alpha$ dimensions
is \citep{Tarasov2015,Tarasov2006Chaos,Tarasov2005AnnPhys}
\begin{equation}
\int_{|\vec{k}|\leq k}d^{\alpha}k=\frac{2\pi^{\alpha/2}}{\Gamma(\alpha/2)}\cdot\int_{0}^{k}k'{}^{\alpha-1}dk'=\frac{2\pi^{\alpha/2}}{\Gamma(\alpha/2)}\cdot\frac{k^{\alpha}}{\alpha}.\label{eq:Volume of a Ball}
\end{equation}

Therefore, the cumulative number of states becomes 
\begin{equation}
G(E)=\frac{1}{(2\pi)^{\alpha}}\cdot\frac{2\pi^{\alpha/2}}{\Gamma(\alpha/2)}\cdot\frac{1}{\alpha}\cdot k(E)^{\alpha},
\end{equation}
where the factor of $2$ accounts for spin degeneracy.

\paragraph{Origin and combination of the $(2\pi)$ factors.}

The factors of $(2\pi)$ appearing in the expressions for $G(E)$
and $g(E)$ (below) have two distinct origins: (i) the standard phase-space
normalization $(2\pi)^{-\alpha}$ (one factor of $2\pi$ per dimension
in $\alpha$ dimensions), and (ii) the hyperspherical surface factor
$S_{\alpha-1}=2\pi^{\alpha/2}/\Gamma(\alpha/2)$ coming from the reduction
of the $\alpha$-dimensional integral to radial coordinates in non-integer
dimension. Thus, schematically, 
\[
G(E)\;\propto\;\frac{2}{(2\pi)^{\alpha}}\int d^{\alpha}k\;=\;\frac{2}{(2\pi)^{\alpha}}S_{\alpha-1}\int_{0}^{k(E)}k^{\alpha-1}dk,
\]
so that all $(2\pi)$ factors could in principle be absorbed into
a single geometric prefactor. In the present work we prefer to keep
the $(2\pi)$ factors explicit, in order to make transparent the standard
phase-space normalization and the separate geometric contribution
from the non-integer-dimensional hyperspherical surface. %================== END PATCH M2 ==================

To express $k$ as a function of energy, we use the free-particle
dispersion relation: 
\begin{equation}
E=E_{0}+\frac{\hbar^{2}k^{2}}{2m}\quad\Rightarrow\quad k(E)=\sqrt{\frac{2m}{\hbar^{2}}(E-E_{0})},
\end{equation}
where $E_{0}$ is the minimum (threshold) energy. Thus, 
\begin{equation}
k^{\alpha}=\left(\frac{2m}{\hbar^{2}}\right)^{\alpha/2}(E-E_{0})^{\alpha/2}.
\end{equation}

Substituting into the expression for $G(E)$, we obtain: 
\begin{equation}
G(E)=\frac{1}{(2\pi)^{\alpha}}\cdot\frac{2\pi^{\alpha/2}}{\Gamma(\alpha/2)}\cdot\frac{1}{\alpha}\cdot\left(\frac{2m}{\hbar^{2}}\right)^{\alpha/2}\cdot(E-E_{0})^{\alpha/2}.
\end{equation}

Grouping constants and using the abbreviation, we define: 
\begin{equation}
C_{\alpha}=\frac{1}{(2\pi)^{\alpha}}\cdot\frac{2\pi^{\alpha/2}}{\Gamma(\alpha/2)}\cdot\frac{1}{\alpha}\cdot\left(\frac{2m}{\hbar^{2}}\right)^{\alpha/2},
\end{equation}
we arrive at the expression\footnote{\begin{flushleft}
In He (1990), the symbol $G_{e}(E)$ denotes the electronic DOS. In
the present work, we adopt the notation $g_{q}(E)$ to clearly emphasize
its origin from the application of the $q$-deformed derivative operator
to the cumulative number of states, $G(E)$.
\par\end{flushleft}} 
\begin{equation}
G(E)=C_{\alpha}(E-E_{0})^{\alpha/2}.\label{eq:G(E)-Fractional}
\end{equation}

\subsection{Electronic density of states}

From the cumulative distribution of states, Eq. (\ref{eq:G(E)-Fractional}),
we obtain the undeformed electronic DOS, in a non-integer-dimensional
space, by differentiating: 
\begin{equation}
g(E)=\frac{dG}{dE}=C_{\alpha}\cdot\frac{\alpha}{2}(E-E_{0})^{\frac{\alpha}{2}-1}.
\end{equation}
Substituting the explicit form of $C_{\alpha}$, we get: 
\begin{equation}
g(E)=\frac{1}{(2\pi)^{\alpha}}\cdot\frac{2\pi^{\alpha/2}}{\Gamma(\alpha/2)}\cdot\frac{1}{2}\cdot\left(\frac{2m}{\hbar^{2}}\right)^{\alpha/2}\cdot(E-E_{0})^{\frac{\alpha}{2}-1}.
\end{equation}

Now, instead of the standard derivation, we apply the $q-$deformed
derivative, Eq. (\ref{eq:q-Deformed-Derivative}) to $G(E).$ Following
this new approach, we obtain
\begin{equation}
g_{q}(E)=\mathcal{D}_{q}[G](E)=C_{\alpha}\cdot\frac{\alpha}{2}(E-E_{0})^{\frac{\alpha}{2}-1}\cdot\left[1+(1-q)(E-E_{0})\right].\label{eq:dos_q_final}
\end{equation}
Hence, the full expression for the DOS becomes 
\begin{equation}
g_{q}(E)=\frac{1}{(2\pi)^{\alpha}}\cdot\frac{2\pi^{\alpha/2}}{\Gamma(\alpha/2)}\cdot\frac{1}{2}\cdot\left(\frac{2m}{\hbar^{2}}\right)^{\alpha/2}\cdot(E-E_{0})^{\frac{\alpha}{2}-1}\cdot\left[1+(1-q)(E-E_{0})\right].
\end{equation}

This rigorous derivation is built up without any empirical correction
factors, clearly establishing the deformation as naturally arising
from the mathematical structure of the generalized derivative.

\begin{figure}[h!]
\centering \includegraphics[width=0.75\textwidth]{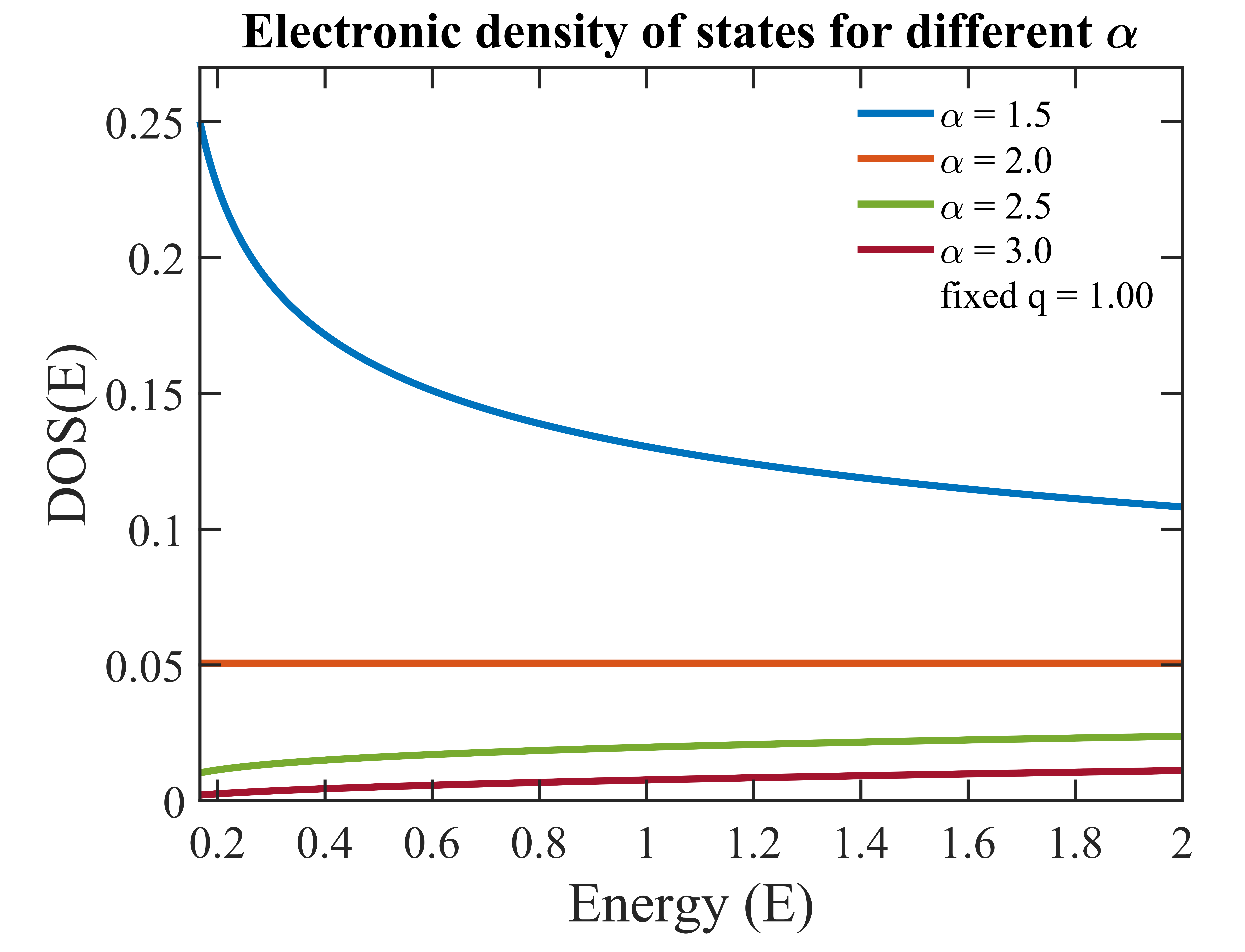}
\caption{\label{fig:DOS_alpha}Electronic DOS for different values of the fractal
dimension $\alpha$. The literal DOS curves are plotted slightly away
from the threshold $E_{0}$ to avoid the integrable near-edge divergence
from dominating the linear-scale visualization.}
\end{figure}

As seen from Figure \ref{fig:DOS_alpha}, depending on the magnitude
of $\alpha$ both concave and convex forms are possible. Also, the
Figure \ref{fig:DOS_alpha} illustrates the electronic DOS as a function
of energy for different values of the non-integer dimension parameter
$\alpha$. The inferred dimension $\alpha<3$ implies reduced phase-space
accessibility, consistent with phonon scattering suppression and dimensional
crossover phenomena, commonly observed in low-dimensional or anisotropic
solids \citep{he1990,Alexander1982,nakayama1994dynamical,Balandin2011,Hu2015,Bottani2018,Sadeghi2013}.
This behavior is consistent with the geometric interpretation of Hausdorff
dimension applied to anisotropic or fractal media, as proposed in
He\textquoteright s model \citep{he1990}.

\subsection{Recovery of He's original result ($q\to1$ limit)}

\label{sec:recovery_he}

In the limit $q\to1$, the $q$-deformed derivative $D_{q}$ reduces
to the ordinary derivative. Therefore, the deformed DOS \eqref{eq:dos_q_final}
reduces to the standard non-integer-dimensional DOS 
\begin{equation}
g_{q=1}(E)=\frac{dG(E)}{dE}=\frac{2}{(2\pi)^{\alpha}}\,\frac{S_{\alpha-1}}{\alpha}\,\frac{d}{dE}\Big[k(E)^{\alpha}\Big]\;\propto\;(E-E_{0})^{\alpha/2-1},\label{eq:dos_q1}
\end{equation}
with $S_{\alpha-1}=2\pi^{\alpha/2}/\Gamma(\alpha/2)$ and the free-particle
dispersion $k(E)=\sqrt{2m(E-E_{0})}/\hbar$. This is precisely the
same scaling and prefactor structure obtained in He's fraction-dimensional
mapping of anisotropic solids to an isotropic $\alpha$-dimensional
system \citep{he1990}. We thus keep the notation $g_{q=1}(E)$ and
no additional volume-like quantity $V_{\alpha}$ is required: all
geometric information is already encoded in the Hausdorff measure
and the normalization $(2\pi)^{-\alpha}$. %================== END PATCH M3 ==================

\section{Phonon Density of States in Non-Integer-Dimensional Space}

We now consider the number of vibrational (phonon) modes with frequency
less than or equal to $\omega$, but in a non-integer-dimensional
momentum space of Hausdorff dimension $\alpha$.

The number of modes is then given by 
\begin{equation}
G_{p}(\omega)=\frac{1}{(2\pi)^{\alpha}}\int_{|\vec{k}|\leq k(\omega)}d^{\alpha}k,
\end{equation}
where $k(\omega)=\omega/v_{s}$, and $v_{s}$ is the speed of sound.

Using the volume \ref{eq:Volume of a Ball} of a ball in $\alpha$
dimensions, we obtain 
\begin{equation}
G_{p}(\omega)=\frac{1}{(2\pi)^{\alpha}}\cdot\frac{2\pi^{\alpha/2}}{\Gamma(\alpha/2)}\cdot\frac{1}{\alpha}\cdot\left(\frac{\omega}{v_{s}}\right)^{\alpha}.\label{eq:Gp(w)}
\end{equation}

The phonon density of states is given by the derivative of this result,
i.e.,
\begin{equation}
g_{p}(\omega)=\frac{dG_{p}}{d\omega}=\frac{1}{(2\pi)^{\alpha}}\cdot\frac{2\pi^{\alpha/2}}{\Gamma(\alpha/2)}\cdot\left(\frac{1}{v_{s}}\right)^{\alpha}\cdot\omega^{\alpha-1}.\label{eq:g_=00007Bp=00007D(=00005Comega)}
\end{equation}

Equation \ref{eq:g_=00007Bp=00007D(=00005Comega)} includes the classical
Debye form in the limit $\alpha=3$. For general $\alpha$, it defines
the spectral distribution of phonon modes in fractal media.

\begin{figure}[h!]
\centering \includegraphics[width=0.75\textwidth]{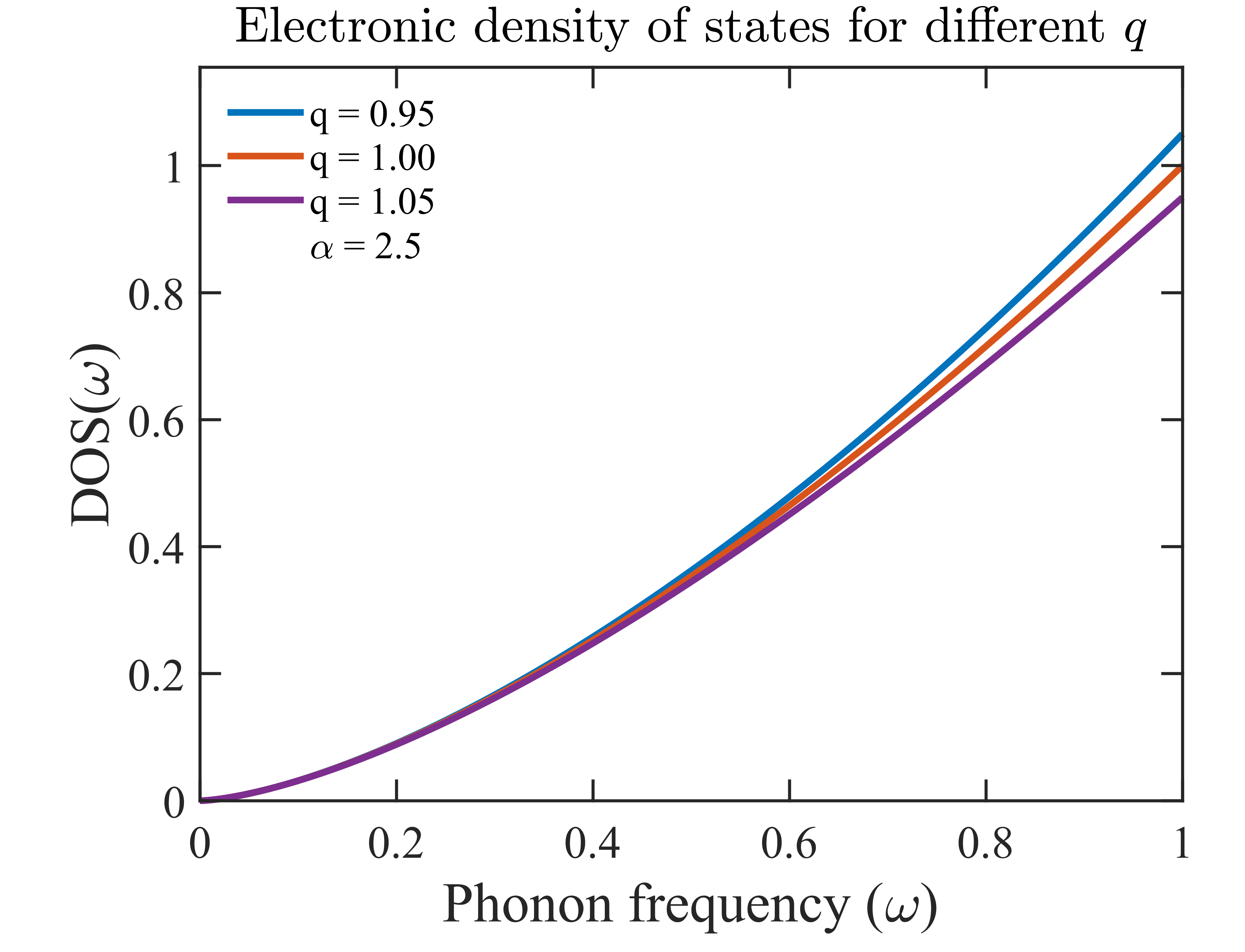}
\caption{\label{fig:DOS_phonon_q}Phonon DOS for different $q$-deformations.
$\alpha=2.5$}
\end{figure}

Figure \ref{fig:DOS_phonon_q} presents the phonon DOS modified by
the $q$-deformed framework. As $q$ deviates from $1$, the low-energy
phonon contributions are enhanced or suppressed, introducing non-trivial
thermodynamic behavior. This deformation is associated with systems
exhibiting correlations, constraints, or generalized statistics in
the Tsallis formalism \citep{tsallis1988}.

%==================== PATCH M4 (after the Fig. 2 discussion paragraph) ====================
%==================== PATCH M4 (insert after the paragraph on Fig. 2, before Subsection A) ====================

\paragraph{Low-frequency behavior and log-scale representation (Fig.~\eqref{fig:DOS_phonon_q}).}

The $q$-deformed phonon DOS derived below, see Eq.~(\ref{eq:gpq_qdeformed}),
has the general structure 
\begin{equation}
g_{p,q}(\omega)\;\propto\;\omega^{\alpha-1}\,\bigl[1+(1-q)\,\omega\bigr],
\end{equation}
with the fractal Hausdorff dimension $\alpha$ and the deformation
parameter $q$. For $\omega\to0$ we have 
\begin{equation}
1+(1-q)\,\omega\;=\;1+\mathcal{O}(\omega),
\end{equation}
so that the leading low-frequency behavior is 
\begin{equation}
g_{p,q}(\omega)\sim\omega^{\alpha-1},\qquad(\omega\to0),
\end{equation}
\emph{independent of $q$} to leading order. On a log-log plot this
implies that the low-frequency slope is the same for all $q$ and
is equal to $(\alpha-1)$, while the $q$-dependence only produces
subleading corrections that become visible at larger $\omega$ (or,
equivalently, at higher temperatures once folded with the Bose factor).
Therefore, the low-$\omega$ part of the curves in Figure~\eqref{fig:DOS_phonon_q}
remains almost unchanged when $q$ is varied, and the main effect
of the deformation appears in the intermediate-to-high frequency regime.
In other words, the low-$\omega$ part is expected to remain almost
identical across $q$ in a log/log-log representation, consistent
with the interpretation that $q$ primarily controls higher-energy
deviations from the Debye-like spectrum. %================== END PATCH M4 ====================

\subsection{Specific heat: derivation of equation's standard Debye approach}

We evaluate the internal energy associated with phonon modes in a
medium of fractional Hausdorff dimension $\alpha$: 
\begin{equation}
U(T)=\int_{0}^{\omega_{D}}\hbar\omega\cdot g_{p}(\omega)\cdot n_{B}(\omega,T)\,d\omega,
\end{equation}
where $g_{p}(\omega)$ is the phonon DOS, $n_{B}(\omega,T)=1/(e^{\hbar\omega/k_{B}T}-1)$
is the Bose-Einstein distribution, and $\omega_{D}$ is the (temperature-independent)
Debye cutoff frequency.

In the previous section, we derived the DOS in non-integer-dimensional
space, given by Eq. (\ref{eq:g_=00007Bp=00007D(=00005Comega)}) 
\begin{equation}
g_{p}(\omega)=\frac{1}{(2\pi)^{\alpha}}\cdot\frac{2\pi^{\alpha/2}}{\Gamma(\alpha/2)}\cdot\left(\frac{1}{v_{s}}\right)^{\alpha}\cdot\omega^{\alpha-1},
\end{equation}
where $v_{s}$ is the speed of sound.

We change variables to $x=\hbar\omega/k_{B}T$, so $\omega=\frac{k_{B}T}{\hbar}x$
and $d\omega=\frac{k_{B}T}{\hbar}dx$. The internal energy becomes
\begin{align}
U(T) & =\frac{1}{(2\pi)^{\alpha}}\cdot\frac{2\pi^{\alpha/2}}{\Gamma(\alpha/2)}\cdot\left(\frac{1}{v_{s}}\right)^{\alpha}\cdot\int_{0}^{\omega_{D}}\frac{\hbar\omega^{\alpha}}{e^{\hbar\omega/k_{B}T}-1}\,d\omega\\
 & =\left(\frac{k_{B}T}{\hbar}\right)^{\alpha+1}\cdot\frac{1}{(2\pi)^{\alpha}}\cdot\frac{2\pi^{\alpha/2}}{\Gamma(\alpha/2)}\cdot\left(\frac{1}{v_{s}}\right)^{\alpha}\cdot\hbar\cdot\int_{0}^{T_{D}/T}\frac{x^{\alpha}}{e^{x}-1}dx,
\end{align}
where $T_{D}=\hbar\omega_{D}/k_{B}$ is the Debye temperature.

In the low-temperature limit $T\ll T_{D}$, the upper limit of the
integral tends to infinity and the integral can be evalueted as \citep{Abramowitz1964,Gradshteyn2007}
\begin{equation}
\int_{0}^{\infty}\frac{x^{\alpha}}{e^{x}-1}dx=\Gamma(\alpha+1)\zeta(\alpha+1),
\end{equation}
where $\zeta(s)$ is the Riemann zeta function.

Therefore, the internal energy of the specific heat scale as 
\begin{equation}
U(T)\propto T^{\alpha+1}\quad\Rightarrow\quad C_{v}(T)=\frac{dU}{dT}\propto T^{\alpha}.
\end{equation}

This confirms that in a space with Hausdorff dimension $\alpha$,
the specific heat scales with temperature as $T^{\alpha}$ at low
temperatures. 

\subsection{Specific heat with deformed phonon density of states in a non-integer
dimensional space}

To include non-linear effects in the phonon statistics, we apply the
$q$-deformed derivative operator, Eq.(\ref{eq:q-Deformed-Derivative}),
to the cumulative phonon mode distribution $G_{p}(\omega)$ (the undeformed
expression) derived in Eq. (\ref{eq:Gp(w)}).

The $q$-deformed phonon DOS becomes 
\begin{equation}
g_{p,q}(\omega):=\mathcal{D}_{q}[G_{p}](\omega)=\left[1+(1-q)\omega\right]\cdot\frac{dG_{p}}{d\omega}.
\end{equation}

Computing the derivative, 
\begin{equation}
\frac{dG_{p}}{d\omega}=\frac{1}{(2\pi)^{\alpha}}\cdot\frac{2\pi^{\alpha/2}}{\Gamma(\alpha/2)}\cdot\left(\frac{1}{v_{s}}\right)^{\alpha}\cdot\omega^{\alpha-1},
\end{equation}
we obtain 
\begin{equation}
g_{p,q}(\omega)=\frac{1}{(2\pi)^{\alpha}}\cdot\frac{2\pi^{\alpha/2}}{\Gamma(\alpha/2)}\cdot\left(\frac{1}{v_{s}}\right)^{\alpha}\cdot\omega^{\alpha-1}\cdot\left[1+(1-q)\omega\right].\label{eq:gpq_qdeformed}
\end{equation}

Now, the internal energy becomes 
\begin{equation}
U_{q}(T)=\int_{0}^{\omega_{D}}\hbar\omega\cdot g_{p,q}(\omega)\cdot\frac{1}{e^{\hbar\omega/k_{B}T}-1}d\omega.
\end{equation}

Using the change of variables $x=\hbar\omega/k_{B}T$ and $\omega=\frac{k_{B}T}{\hbar}x$,
we obtain 
\begin{align}
U_{q}(T) & =\frac{1}{(2\pi)^{\alpha}}\cdot\frac{2\pi^{\alpha/2}}{\Gamma(\alpha/2)}\cdot\left(\frac{1}{v_{s}}\right)^{\alpha}\cdot\left(\frac{k_{B}T}{\hbar}\right)^{\alpha+1}\cdot\hbar\nonumber \\
 & \quad\cdot\int_{0}^{T_{D}/T}\frac{x^{\alpha}\left[1+(1-q)\frac{k_{B}T}{\hbar}x\right]}{e^{x}-1}dx,
\end{align}
where $T_{D}=\hbar\omega_{D}/k_{B}$ is the Debye temperature.

Separating terms inside the integral: 
\begin{equation}
U_{q}(T)=AT^{\alpha+1}\left[\int_{0}^{\infty}\frac{x^{\alpha}}{e^{x}-1}dx+(1-q)\cdot\frac{k_{B}T}{\hbar}\cdot\int_{0}^{\infty}\frac{x^{\alpha+1}}{e^{x}-1}dx\right],
\end{equation}
where the prefactor $A$ is 
\begin{equation}
A=\frac{1}{(2\pi)^{\alpha}}\cdot\frac{2\pi^{\alpha/2}}{\Gamma(\alpha/2)}\cdot\left(\frac{1}{v_{s}}\right)^{\alpha}\cdot\hbar\cdot\left(\frac{k_{B}}{\hbar}\right)^{\alpha+1}\label{eq:prefactor A}
\end{equation}
and using the standard integral 
\[
\int_{0}^{\infty}\frac{x^{\mu}}{e^{x}-1}dx=\Gamma(\mu+1)\zeta(\mu+1),
\]
we obtain for the internal energy 
\begin{equation}
U_{q}(T)=AT^{\alpha+1}\left[\Gamma(\alpha+1)\zeta(\alpha+1)+(1-q)\cdot\frac{k_{B}T}{\hbar}\cdot\Gamma(\alpha+2)\zeta(\alpha+2)\right].
\end{equation}

Differentiating with respect to $T$, the specific heat becomes 
\begin{align}
C_{V,q}(T) & =\frac{dU_{q}}{dT}\nonumber \\
 & =A(\alpha+1)\Gamma(\alpha+1)\zeta(\alpha+1)T^{\alpha}\nonumber \\
 & \quad+A(1-q)\cdot\frac{k_{B}}{\hbar}\left[(\alpha+2)\Gamma(\alpha+2)\zeta(\alpha+2)T^{\alpha+1}\right].\label{eq:Cv,q-1}
\end{align}

In the limit $q\to1$, Eq. (\ref{eq:Cv,q-1}) reduces to the undeformed
fractional-dimensional Debye form

\begin{equation}
C_{V}^{(D,\alpha)}(T):=\lim_{q\to1}C_{V,q}(T)=A(\alpha+1)\Gamma(\alpha+1)\zeta(\alpha+1)\,T^{\alpha}\equiv A_{1}T^{\alpha}.\label{eq:debye_limit_fractional}
\end{equation}

Therefore, the $q$-deformed specific heat can be written as

\begin{equation}
C_{V,q}(T)=C_{V}^{(D,\alpha)}(T)+A_{2}(1-q)T^{\alpha+1},\label{eq:cvq_debye_plus_correction}
\end{equation}

showing explicitly that the deformation introduces a temperature-dependent
correction governed by $(1-q)T^{\alpha+1}$.

\subsection{Factorization of the deformed specific heat expression}

From the full derivation of the deformed phonon specific heat, we
obtained Eq. (\ref{eq:Cv,q-1}) along with the prefactor (\ref{eq:prefactor A}).

Let us define the constants 
\begin{align}
A_{1} & :=A(\alpha+1)\Gamma(\alpha+1)\zeta(\alpha+1),\label{A1}\\
A_{2} & :=A\cdot\frac{k_{B}}{\hbar}(\alpha+2)\Gamma(\alpha+2)\zeta(\alpha+2).\label{A2}
\end{align}

Then the specific heat can be written in the form 
\begin{equation}
C_{V,q}(T)=A_{1}T^{\alpha}+A_{2}(1-q)T^{\alpha+1},
\end{equation}
which can also be factorized as 
\begin{equation}
C_{V,q}(T)=T^{\alpha}\left[A_{1}+A_{2}(1-q)T\right].\label{eq:Cvq-Phonons}
\end{equation}

It is important to note that this expression has physically meaningful
prefactors, derived from the deformed Debye model in non-integer-dimensional
space.

\subsection{Debye saturation behavior and the role of the cutoff term}

In real crystalline solids, the specific heat exhibits a well-known
saturation behavior as the temperature approaches or exceeds the Debye
temperature $T_{D}$ \citep{Kittel2005}. This arises from the complete
thermal excitation of all phonon modes in the system.

At high temperatures, $T\gg T_{D}$, all vibrational modes are populated,
and the heat capacity asymptotically approaches the classical Dulong-Petit
limit 
\[
C_{V}(T)\rightarrow3R
\]
for monatomic solids, where where $R$ is the gas constant.

However, the result (\ref{eq:Cvq-Phonons}) for the deformed heat
capacity can grow unboundedly for $\alpha>1$ as $T\rightarrow\infty$,
potentially diverging or overshooting the physical behavior.

To incorporate the observed saturation of heat capacity at high temperatures,
we include a cutoff factor in our deformed model, of the form \citep{weberszpil2025microscopic}
\begin{equation}
\frac{1}{1+\left(\frac{T}{T_{D}}\right)^{n}}\label{eq:Cutoff}
\end{equation}
This factor mimics the asymptotic behavior predicted by Debye theory,
where the specific heat approaches a finite value as all phonon modes
are excited. For $T\ll T_{D}$, the correction is negligible, preserving
the low-temperature power-law scaling, while for $T\gg T_{D}$, the
term (\ref{eq:Cutoff}) suppresses the divergence and reflects the
physical saturation of $C_{V}(T)$. The exponent $n$ controls the
sharpness of this transition and can be adjusted to match the observed
thermal behavior of the material. Concretely, for $T\ll T_{D}$, 
\begin{equation}
\left(\frac{T}{T_{D}}\right)^{n}\ll1\quad\Rightarrow\quad\left(1+\left(\frac{T}{T_{D}}\right)^{n}\right)^{-1}\approx1,
\end{equation}
meaning the factor has negligible effect in the low-temperature regime.

For $T\gg T_{D}$, 
\begin{equation}
\left(\frac{T}{T_{D}}\right)^{n}\gg1\quad\Rightarrow\quad\left(1+\left(\frac{T}{T_{D}}\right)^{n}\right)^{-1}\rightarrow0,
\end{equation}
meaning the specific heat growth is suppressed and saturates. 

\subsubsection{Physical constraints on the saturation exponent $n$}

The saturation exponent $n$ in Eq. (\eqref{eq:Cutoff}) is not arbitrary
but must satisfy certain physical requirements to ensure thermodynamic
consistency:
\begin{enumerate}
\item \textbf{Positivity:} $n>0$ ensures that the cutoff function monotonically
decreases from 1 to 0 as $T/T_{D}$ increases from 0 to $\infty$.
\item \textbf{Asymptotic behavior:} For $T\ll T_{D}$, we require 
\begin{equation}
\frac{1}{1+(T/T_{D})^{n}}\approx1-(T/T_{D})^{n}+\mathcal{O}[(T/T_{D})^{2n}],
\end{equation}
which introduces only exponentially small corrections at low temperatures,
preserving the $T^{\alpha}$ power-law behavior established experimentally.
\item \textbf{Smoothness of transition:} The value of $n$ controls the
sharpness of the crossover near $T\sim T_{D}$. Within the present
phenomenological cutoff model, smaller values of $n$ correspond to
a smoother and broader crossover, whereas larger values of $n$ produce
a sharper, more Debye-like transition. This interpretation is consistent
with the general expectation that stronger anharmonic phonon-phonon
scattering broadens the effective phonon spectrum, while simpler weakly
bound systems such as rare-gas solids are closer to the harmonic reference
limit \citep{Ramirez2005,RahmanSpangler2002,Klein1976}. In particular: 
\end{enumerate}
\begin{itemize}
\item $n\sim2$: smooth and gradual crossover, appropriate for broader effective
phonon-spectrum cutoffs. Consistent with strongly anharmonic materials
in which phonon-{}-phonon scattering broadens the spectral density
near $T_{D}$ \citep{Ramirez2005,RahmanSpangler2002}; 
\item $n\sim4$--$6$: sharper crossover near $T_{D}$, appropriate for
more well-defined effective Debye cutoffs. Consistent with nearly
harmonic solids such as rare-gas crystals \citep{Klein1976,Ramirez2005}; 
\item $n\to\infty$: Step-function (Heaviside) limit, recovering the ideal
Debye cutoff \citep{Kittel2005}. This case is unphysical but serves
as a useful theoretical benchmark.
\end{itemize}
The numerical ranges above should be understood as phenomenological
guidelines within the present model, not as universal material constants.
\begin{enumerate}
\item \textbf{Relationship to phonon anharmonicity:} Materials with stronger
phonon-phonon anharmonic scattering are expected to exhibit broader
effective phonon spectra near $T_{D}$, which in the present cutoff
model is consistent with smaller values of $n$. By contrast, weakly
bound systems such as rare-gas solids are closer to the harmonic reference
limit and may therefore be associated with sharper cutoffs, i.e.,
larger values of $n$ \citep{Ramirez2005,RahmanSpangler2002,Klein1976}.
The precise relation between $n$ and anharmonicity strength, however,
remains phenomenological.
\item \textbf{Empirical constraint:} As summarized in Table~\ref{tab:Fitted-parameters-for},
the best-fit values satisfy $1.5<n<7$ for all tested systems, with
most cases lying in the range $2<n<4$. In the framework of the present
phenomenological cutoff model, this interval corresponds to moderate
crossover sharpness in the effective phonon cutoff. Any more detailed
interpretation of $n$ in terms of anharmonicity strength remains
phenomenological \citep{Ramirez2005,RahmanSpangler2002,Klein1976}.
\end{enumerate}
We emphasize that $n$ is treated as a phenomenological parameter
extracted from fits to experimental data, analogous to the Debye temperature
$T_{D}$ itself. Future work incorporating explicit anharmonic phonon-phonon
interactions from first principles may provide a microscopic justification
for specific $n$ values in different material classes.

This flexibility allows one to tailor the high-temperature behavior
of the model based on empirical data, ensuring a smooth interpolation
between the low-temperature scaling and the physically correct asymptotic
saturation. 

The final expression, which we refer to as the $\emph{entropic model}$,
is the given by the expression

\begin{equation}
C_{V,q}(T)=T^{\alpha}\left[A_{1}+A_{2}(1-q)\,T\right]\left(1+\left(\frac{T}{T_{D}}\right)^{n}\right)^{-1}.\label{eq:Cv,q}
\end{equation}

The term $\emph{entropic model}$ is employed here to emphasize the
role of the nonextensive entropy parameter $q$, which modulates the
system\textquoteright s deviation from classical extensive thermodynamics.
The presence of the deformation encodes thermodynamic anomalies often
observed in complex or anisotropic materials.

\begin{figure}[h!]
\centering \includegraphics{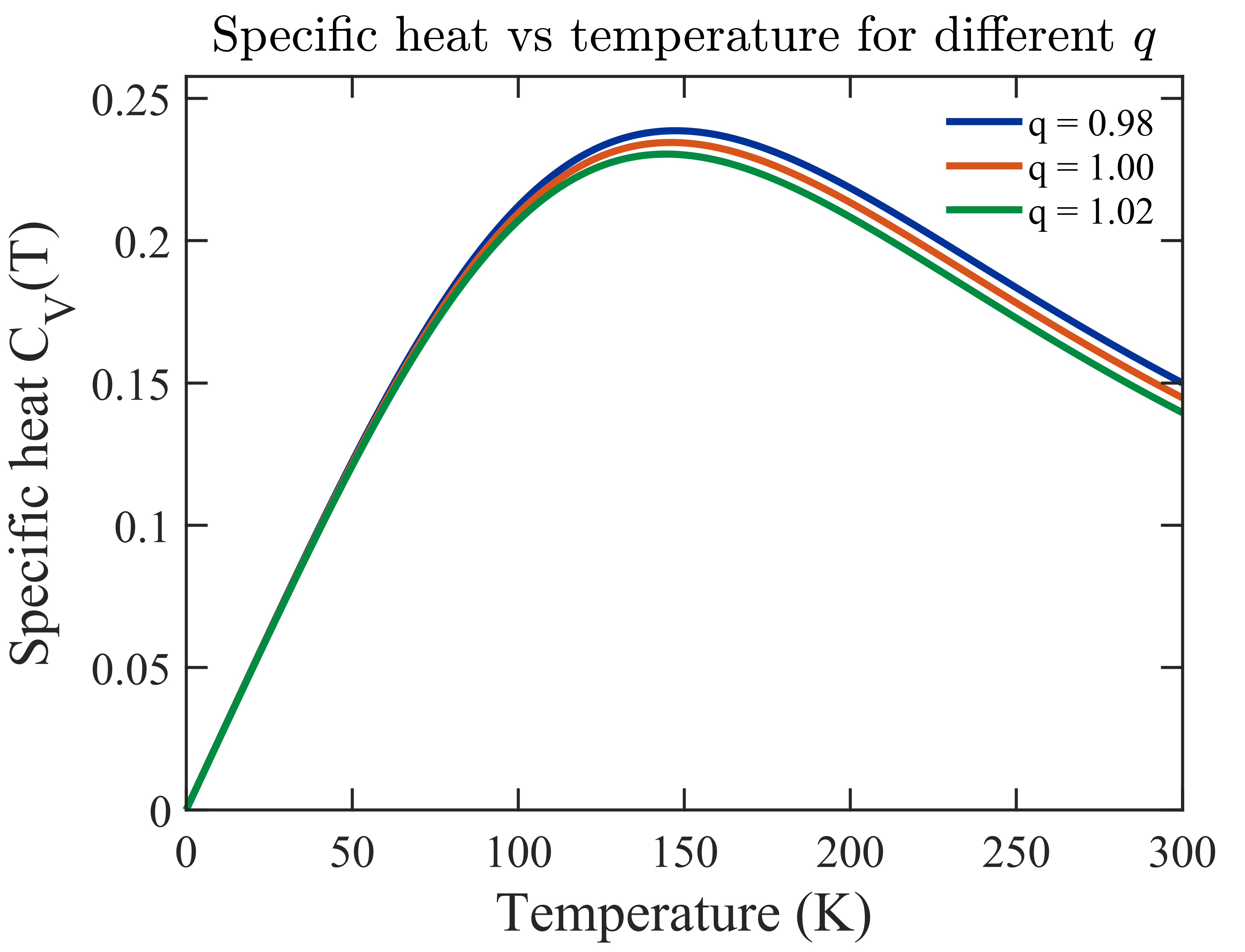}\caption{\label{fig:Specific-heat-Cv}Specific heat $C_{V,q}(T)$ versus temperature
$T$ for different values of $q$, obtained from Eq.(\ref{eq:Cv,q})
with $\alpha=1.0$, $T_{D}=180\,\mathrm{K}$, and $n=2.8$ (with fixed
$A_{1}=2.5\times10^{-3}$ and $A_{2}=1.5\times10^{-5}$).}
\end{figure}

Figure \ref{fig:Specific-heat-Cv} shows the temperature dependence
of the specific heat $C_{V,q}(T)$ for different values of the entropic
parameter $q$, as computed from Eq. (\ref{eq:Cv,q}). For $q=1$,
the model recovers the undeformed Debye-like reference behavior. Small
deviations from $q=1$ produce systematic shifts in the magnitude
and curvature of the heat-capacity curves, especially in the intermediate-to-high
temperature regime, while preserving the low-temperature power-law
scaling controlled by $\alpha$. In this sense, the parameter $q$
may be interpreted as encoding weak nonextensive effects associated
with disorder, heterogeneity, or finite-size constraints in anisotropic
solids. The values of $A_{1}\text{(\ref{A1})}$, $A_{2}$(\ref{A2}),
and $n$ used in Fig. \ref{fig:Specific-heat-Cv} were selected as
representative phenomenological parameters satisfying the thermodynamic
constraints of the model and yielding a physically reasonable crossover
profile. In particular, $A_{1}$ sets the overall amplitude, $A_{2}$
controls the strength of the $q$-dependent correction, and $n$ determines
the smoothness of the saturation near $T_{D}$.

\subsection{Microscopic origin of the $q$--deformation and link to conformable
dynamics}

\label{subsec:micro-origin-q}

Recent developments \citep{weberszpil2025microscopic} show how quenched
disorder and temperature-dependent kinetics produce \emph{emergent
conformable dynamics}. In this framework, a local kinetic coefficient
obeys 
\begin{equation}
\Gamma(\mathbf{r},T)=\Gamma_{0}\,T^{\,1-\mu}\,f(\mathbf{r}),\quad0<\mu<1,\ \ \langle f\rangle=1,\label{eq:Gamma-conformable}
\end{equation}
and the thermal evolution may be written in terms of the \emph{conformable}
(local) derivative of a function $\psi(T)$
\begin{equation}
D_{T}^{(\mu)}\psi:=T^{\,1-\mu}\,\frac{d\psi}{dT},
\end{equation}
which captures memory/heterogeneity without explicit nonlocal kernels.
After coarse-graining, fluctuations of $\Gamma$ generate an effective
memory kernel with a power-law tail, 
\begin{equation}
K(\tau)\sim\tau^{\mu-1},\label{eq:K(tau)}
\end{equation}
consistent with non-Markovian relaxation. Such power-law kernels and
the associated non-Markovian response are standard hallmarks of anomalous
transport and CTRW/fractional kinetics; see the reviews by \citep{MetzlerKlafter2000,MetzlerKlafter2004}.
Here $K(\tau)$ denotes the effective memory kernel, where $\tau=t-t'>0$
is the lag time. It quantifies how strongly past states at time $t'$
contribute to the present dynamics at time $t$ after coarse-graining
over the fluctuations of $\Gamma$. In this representation, the emergence
of a power-law tail (\ref{eq:K(tau)}), indicates algebraically decaying
memory and is therefore consistent with non-Markovian relaxation.

Under broad adiabatic conditions, this kinetic deformation maps to
the entropic deformation used here. A convenient identification is
\begin{equation}
q\;\approx\;1+\frac{1}{\mu}\qquad(q>1),\label{eq:mu-q-map}
\end{equation}
so that $q$ encodes the material's heterogeneity/memory strength
via $\mu$. This turns $q$ from a purely empirical fitting knob into
a \emph{physically interpretable} parameter.

A direct implication of this identification is that an independent
estimate of $\mu$ (e.g., from relaxation spectroscopy or impedance/echo
experiments) provides a \emph{prior} for $q$ via \eqref{eq:mu-q-map},
reducing degeneracies in $C_{V}(T)$ fits.

\section{Validation of the $q-$Deformed Model (Entropic Model) Through Experimental
Data}

In this section, we validate the proposed $q-$deformed model for
the specific heat $C_{V,q}(T)$ in anisotropic solids by testing it
against two experimental datasets from the literature: Sapphire ($\alpha$-Al$_{2}$O$_{3}$)
data taken from \citep{Ditmars1982} and reported by ~\citep{InstrumentsTN8a};
and macroscopic cobalt nanowire composites, as reported by Pradhan
et al.~\citep{pradhan2008specific}.

\subsection*{Justification for using $C_{p}$ data in $C_{V}$ modeling}

Although the entropic model developed here is formally structured
to represent the specific heat at constant volume, $C_{V}(T)$, all
experimental data used in this study, such as for sapphire, quartz,
silicon, copper, and germanium, were obtained under constant pressure
conditions, yielding $C_{p}(T)$ measurements.

However, for crystalline solids in the considered temperature ranges
(typically below $1000\,\mathrm{K}$), the thermodynamic difference
between $C_{p}$ and $C_{V}$ is negligible. This is because the correction
term \citep{Kittel2005,Pathria2011}
\begin{equation}
C_{p}-C_{V}=\frac{TV\alpha^{2}}{\kappa_{T}}
\end{equation}
depends on the thermal expansion coefficient $\alpha$, the isothermal
compressibility $\kappa_{T}$, and the molar volume $V$, all of which
remain small in crystalline solids. As a result, $C_{p}\approx C_{V}$
is a well-established approximation in solid-state physics. Therefore,
throughout this work, we adopt the experimentally measured $C_{p}(T)$
as a practical proxy for $C_{V}(T)$ in all model fitting and validation
procedures. 

\subsection{Fit to Sapphire specific heat data}

\begin{figure}[H]
\centering \includegraphics[width=0.65\textwidth]{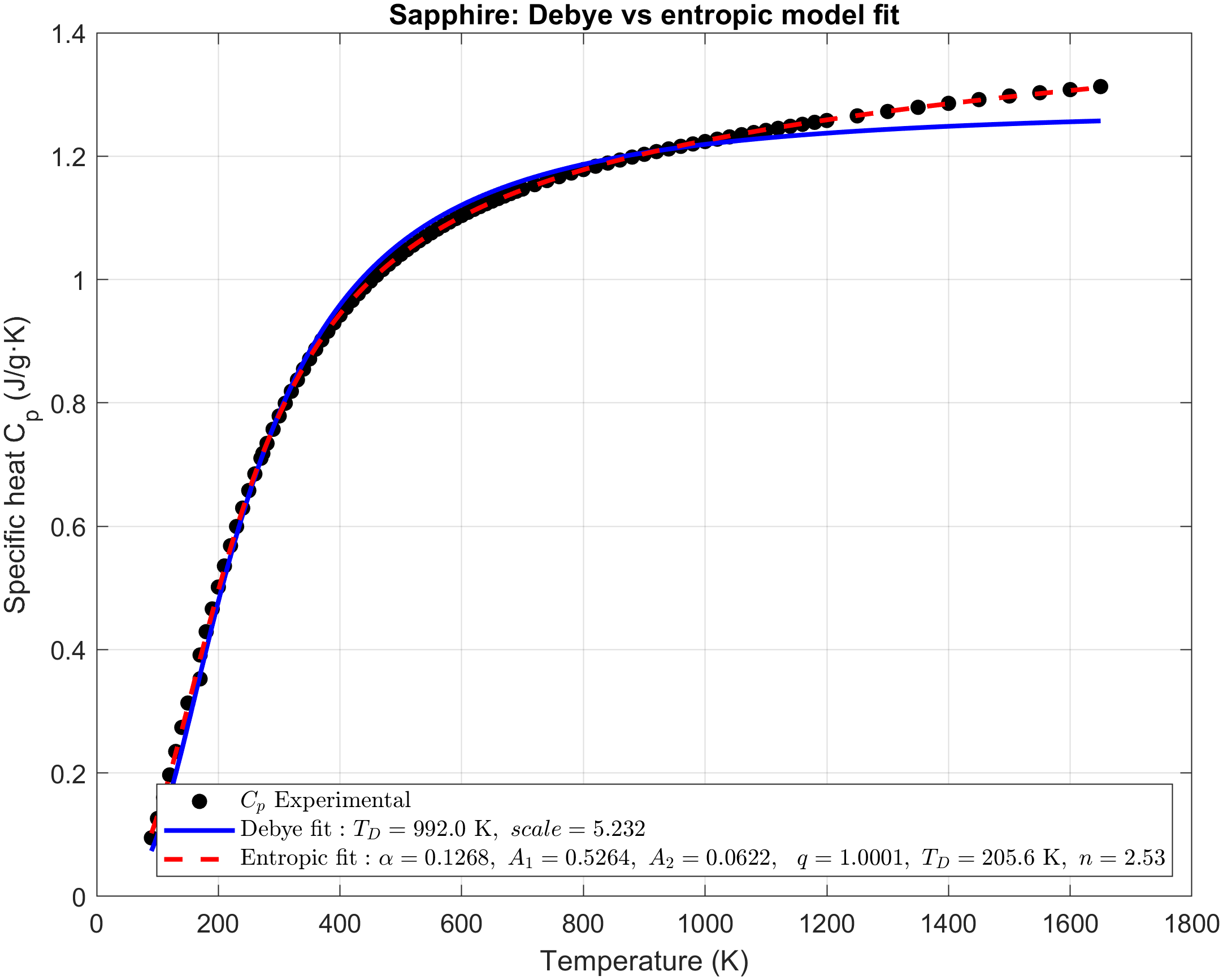}
\caption{\label{fig:Fit-of-the-Sapphire}Best fit of the entropic model, Eq.~(\ref{eq:Cv,q}),
to the specific heat of sapphire ($\alpha$-Al$_{2}$O$_{3}$), yielding
best-fit parameters $\alpha=2.92$, $q=0.979$, $A_{1}=5.00\times10^{-3}$,
and $A_{2}=1.00\times10^{-6}$. }
\end{figure}

The dataset from Ref. \citep{InstrumentsTN8a} covers a wide temperature
range from approximately 90K to 165K. We applied the $q-$deformed
model in a non-integer-dimensional space by fitting the model parameters
(see Fig. \ref{fig:Fit-of-the-Sapphire}). We also plotted in the
same graph a fit to the Debye model. The entropic model provides a
significantly better fit to the experimental data than the classical
Debye approach. Although the entropic model involves a larger number
of adjustable parameters, we emphasize that each parameter carries
a well-defined physical meaning: $\alpha$ encodes the effective Hausdorff
dimensionality of phonon propagation, $q$ quantifies the degree of
nonextensivity, $T_{D}$ sets the thermal scale for mode saturation,
and $n$ controls the sharpness of the high-temperature cutoff.

\subsection{Fit to Cobalt nanowire composite data}

These nanowire composites exhibit strong directional thermal anisotropy,
making them ideal candidates for the deformed derivative model.

\begin{figure}[H]
\centering \includegraphics[width=0.65\textwidth]{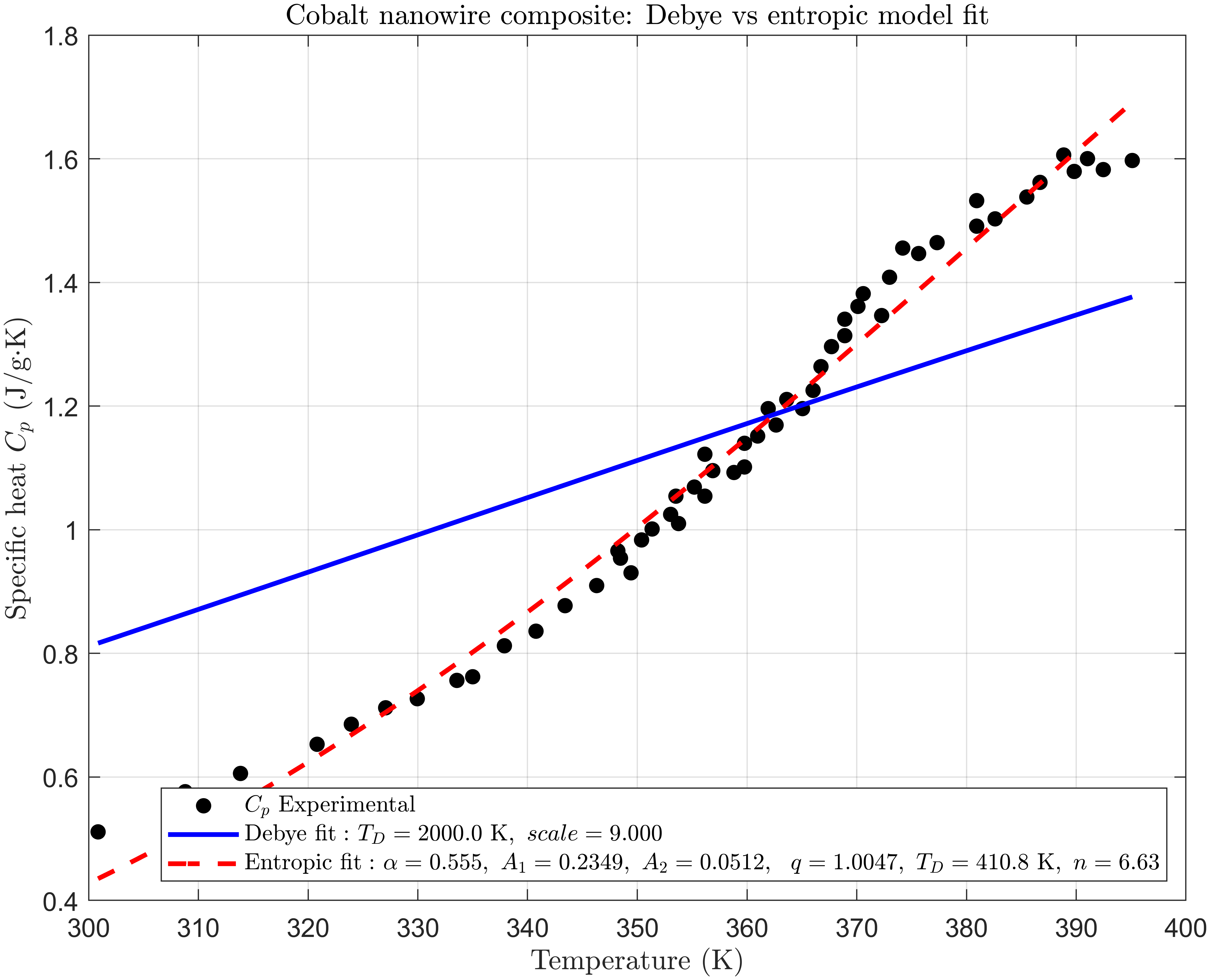}
\caption{\label{fig:Fit-to-cobalt}. Specific heat data for the cobalt nanowire
composite reported by Pradhan et al. \citep{pradhan2008specific},
compared with theoretical fits over the interval $300K$ to $400K$.
The experimental data (black circles) correspond to a cobalt nanowire
array embedded in an anodic aluminum oxide (AAO) matrix, with wire
diameter of approximately $200~nm$. The Debye benchmark fit (blue
solid line), with best-fit parameter $T_{D}\approx1520K$, systematically
underestimates the measured specific heat throughout the investigated
temperature range. In contrast, the entropic saturation model (red
dashed line), with $\alpha=0.555$, $q=1.0047$, $A_{1}=0.2349$,
$A_{2}=0.0512$, $T_{D}=410.8K$, and $n=6.63$, reproduces both the
magnitude and the curvature of the data much more accurately. The
fitted value $\alpha\approx0.56$ suggests an effective reduced dimensionality
associated with anisotropic or confined transport in the nanowire
system, whereas $q\approx1.00$ remains close to the extensive limit.}
\end{figure}

The comparison between the Debye benchmark model and the entropic
saturation model applied to the specific-heat data of the cobalt nanowire
composite (Figure \ref{fig:Fit-to-cobalt}) reveals a clear advantage
of the entropic approach. The dataset extracted from Pradhan et al.
\citep{pradhan2008specific} corresponds to the cobalt nanowire composite/randomly
oriented $Co$ nanowire sample rather than bulk cobalt. Over the measured
interval ($300K$ to $400K)$, the Debye benchmark predicts only a
smooth monotonic increase and significantly underestimates the experimental
data. The corresponding best-fit Debye temperature ($T_{D}\approx1520~\mathrm{K}$)
is unusually high for this effective composite response and produces
a curve that rises too slowly to reproduce the pronounced increase
observed in the data. In contrast, the entropic saturation model captures
both the curvature and the magnitude of the measured $C_{p}(T)$ much
more accurately.

In contrast, the entropic model in Eq.(\ref{eq:Cv,q}) provides excellent
agreement with experimental measurements. The combination of a power-law
growth, a linear-in-$T$ correction, and a saturation cutoff controlled
by $T_{D}$ and $n$ allows the model to reproduce both the curvature
and the leveling behavior of the specific heat. This result suggests
that the entropic formulation, although phenomenological, effectively
captures the complex thermodynamic behavior that goes beyond harmonic
phonon approximations, such as anharmonicity, defect contributions,
and finite-size effects often relevant in nanostructured materials.

\section{Theoretical Framework and Parameter Interpretation}

\subsection{Connection to Tsallis statistical mechanics}

The $q$-deformed derivative employed in this work emerges naturally
from Tsallis statistical mechanics, where the entropy is generalized
to \citep{tsallis1988,Tsallis2009} 
\begin{equation}
S_{q}=k_{B}\frac{1-\sum_{i}p_{i}^{q}}{q-1}
\end{equation}

This generalized entropy leads to modified probability distributions
that exhibit power-law tails and nonextensive behavior. In the context
of phonon systems, this translates to a modified DOS that capture
correlations and memory effects through the $q$-deformation.

The key insight is that anisotropic materials naturally break the
assumptions of extensive thermodynamics. The directional dependence
of thermal transport, combined with microstructural correlations,
creates effective nonlocality that is captured by the $q$-deformed
framework.

\subsection{Parameter identifiability and physical constraints}

To address concerns about parameter identifiability, we impose physical
constraints on the fitting parameters:

\textbf{Hausdorff Dimension $\alpha$:} Constrained to $0<\alpha\leq3$,
reflecting the physical requirement that the effective dimensionality
cannot exceed the embedding space dimension.

\textbf{Deformation Parameter $q$:} Values near unity ($0.95<q<1.05$)
indicate weak nonextensivity, while significant deviations suggest
strong correlations or structural disorder.

\textbf{Saturation Parameters:} The Debye temperature $T_{D}$ must
be physically reasonable (typically $100-2000K$ for most solids),
and the saturation exponent $n$ should be positive to ensure proper
high-temperature behavior.

\section{Other Materials}

\begin{figure}[H]
\centering \includegraphics[width=0.9\textwidth]{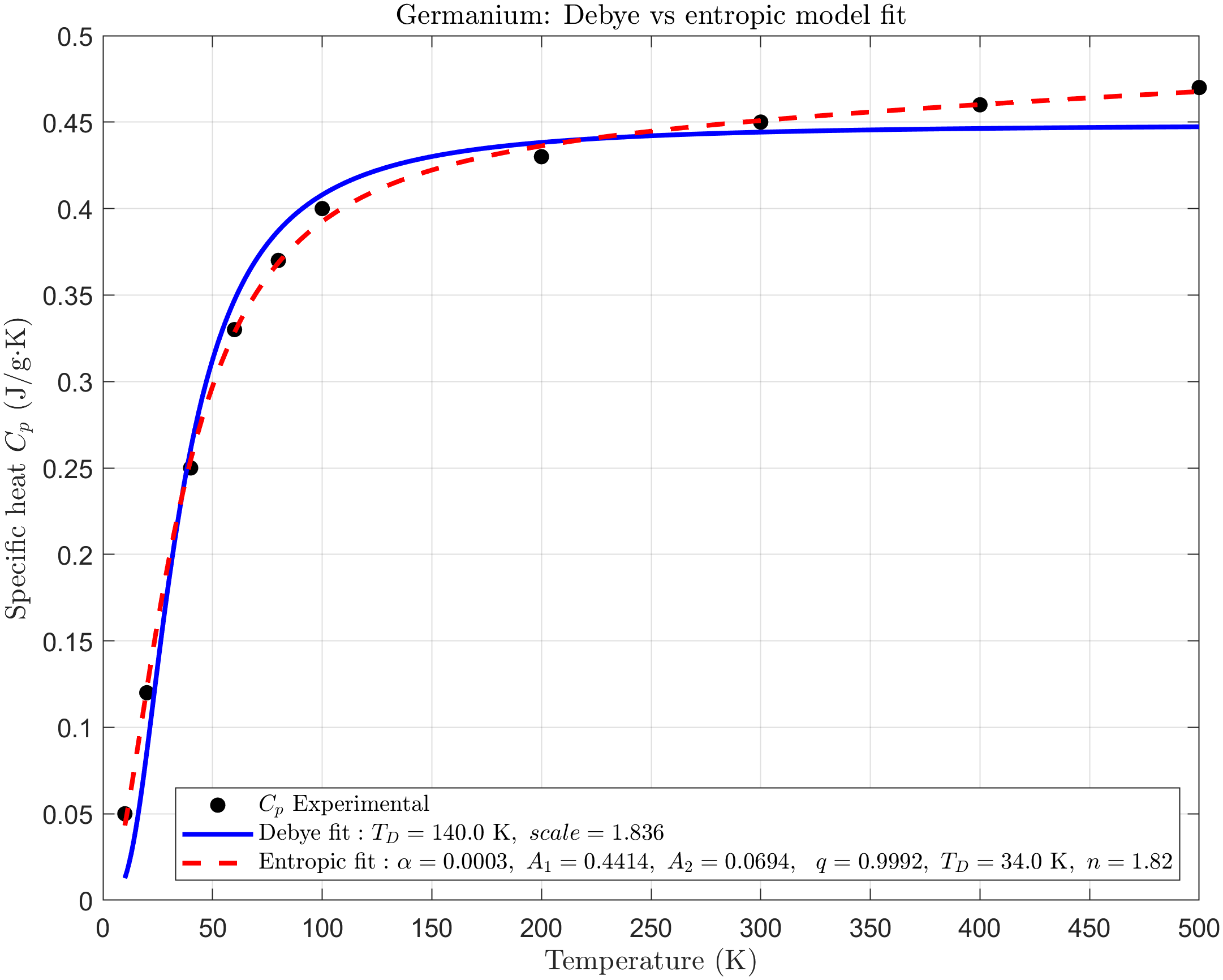}
\caption{\textbf{Specific heat capacity for Germanium:} Comparison between
Debye and entropic saturation model fits.\protect \\
 Data from Piesbergen et al. (1963) \citep{piesbergen1963germanium}.}
\label{fig:germanium_fit}
\end{figure}

\begin{figure}[H]
\centering \includegraphics[width=0.9\textwidth]{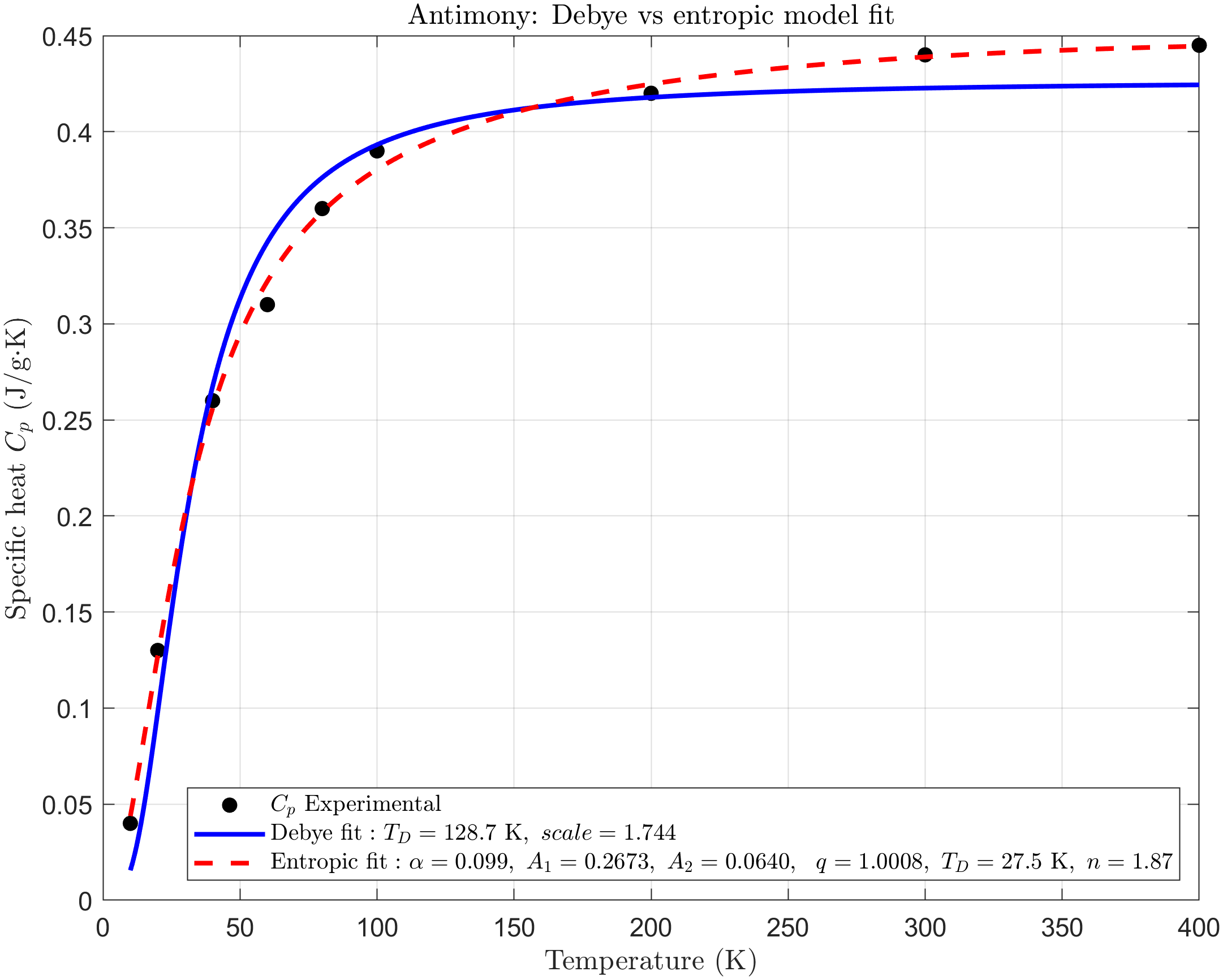}
\caption{\textbf{Specific heat capacity for Antimony:} Fitting comparison between
the Debye model and the modified entropic model.\protect \\
Data from Pradhan et al. (2008) \citep{pradhan2008specific}.}
\label{fig:antimony_fit}
\end{figure}

\begin{figure}[H]
\centering \includegraphics[width=0.9\textwidth]{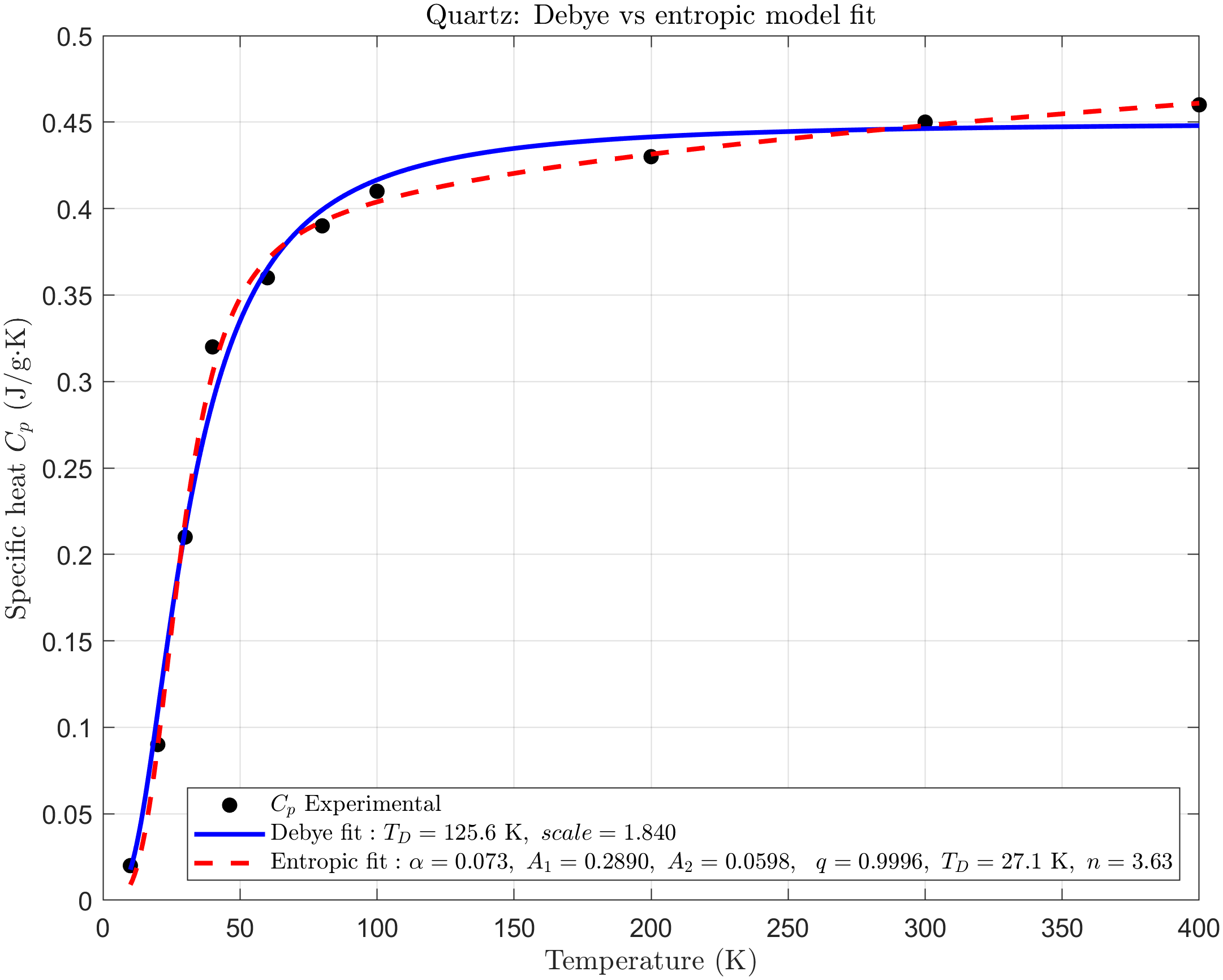}
\caption{\textbf{Specific heat capacity for Quartz (SiO\protect\textsubscript{2}):}
Debye and entropic model comparison.\protect \\
Data from NIST Standard Reference Database \citep{nistshomate}.}
\label{fig:quartz_fit}
\end{figure}

\begin{figure}[H]
\centering \includegraphics[width=0.9\textwidth]{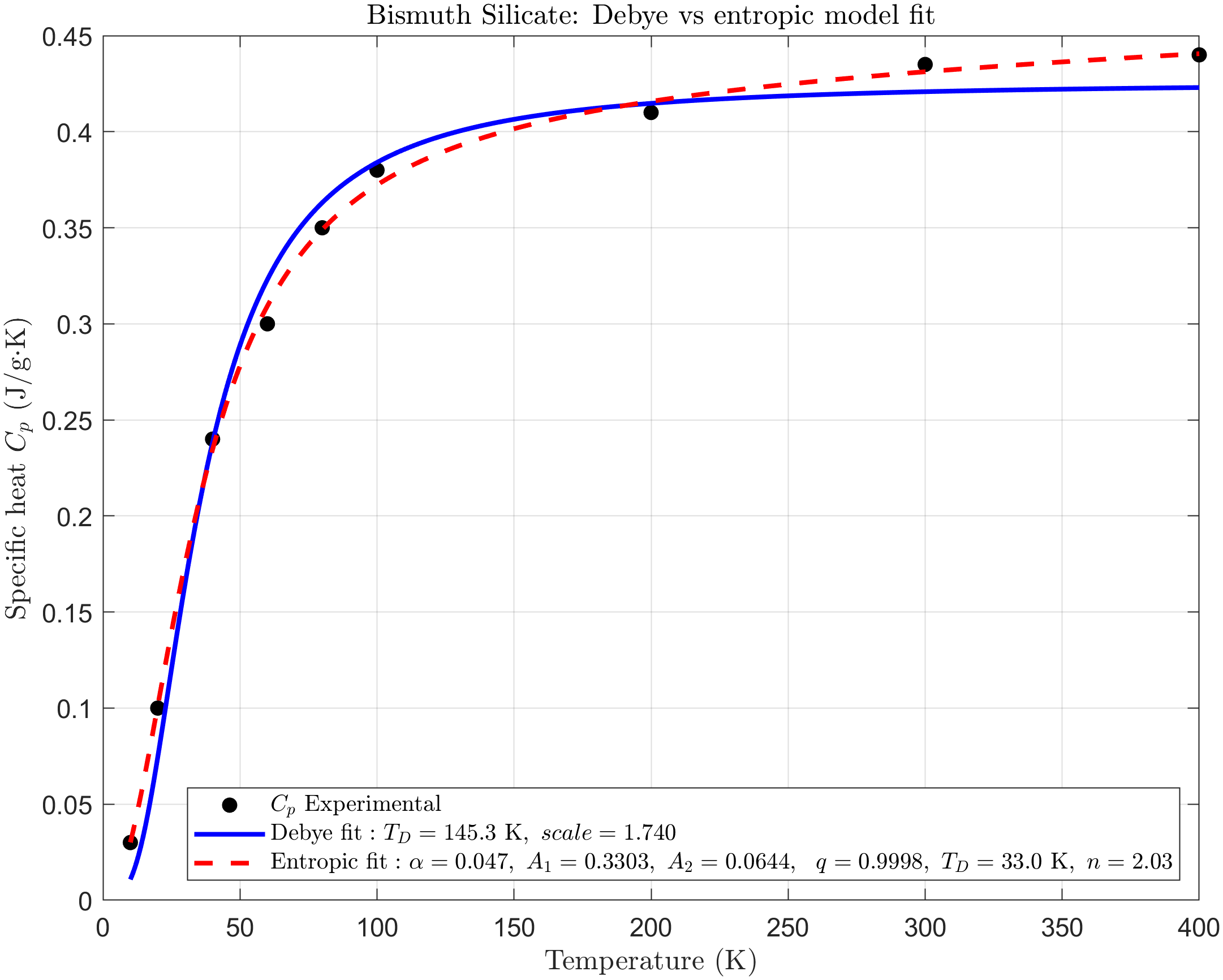}
\caption{\textbf{Specific heat capacity for Bismuth Silicate:} Debye versus
entropic saturation model fit.\protect \\
 Data from Onderka et al. (2015) \citep{onderka2015bismuthsilicate}.}
\label{fig:bismuth_silicate_fit}
\end{figure}

\begin{figure}[H]
\centering \includegraphics[width=0.9\textwidth]{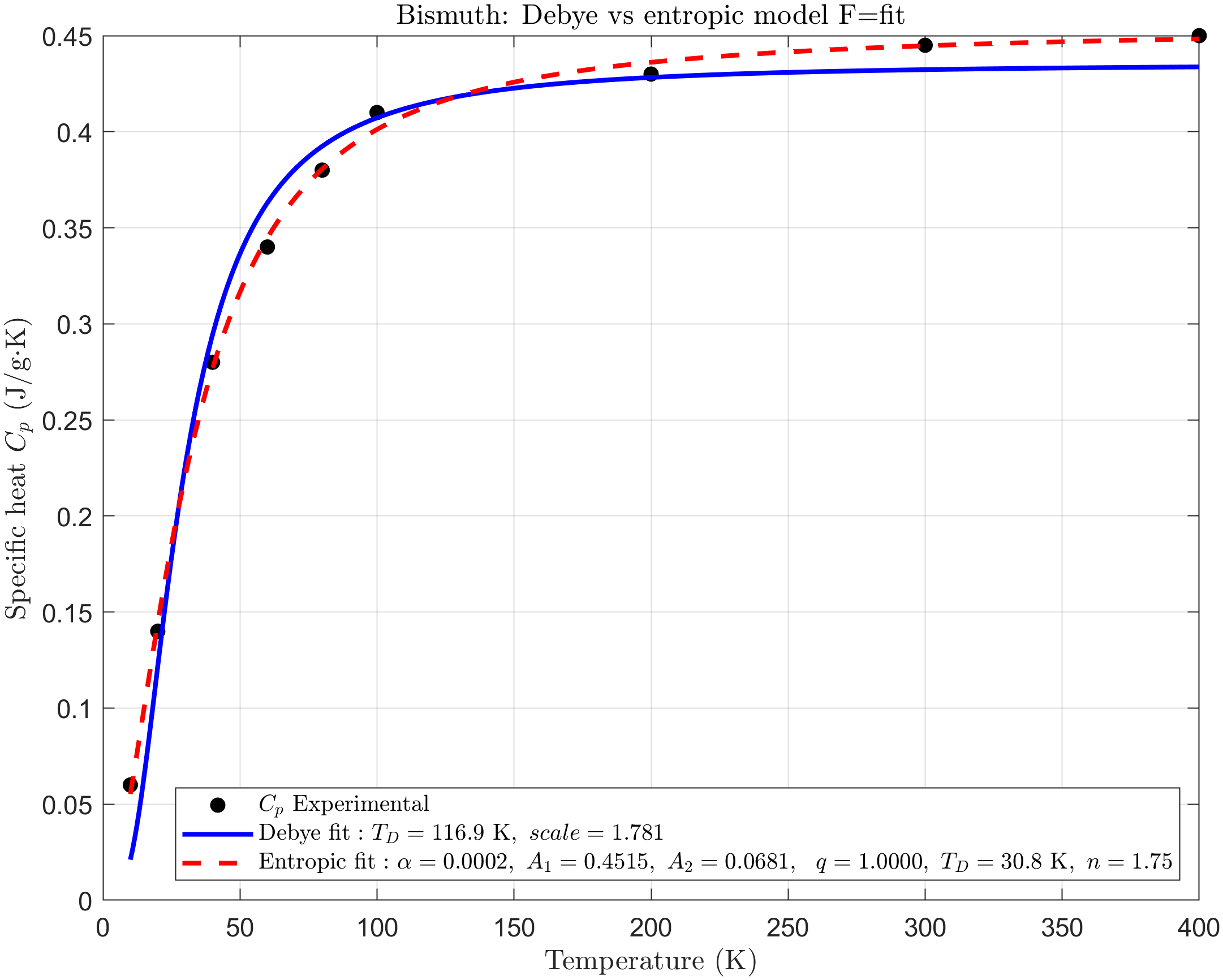}
\caption{\textbf{Specific heat capacity for Bismuth:} Specific heat fit comparison.\protect \\
Data from Pradhan et al. (2008) \citep{pradhan2008specific}.}
\label{fig:bismuth_fit}
\end{figure}

To assess the predictive capability of the modified entropic model
for the specific heat capacity, we additionally performed comparative
fits against the classical Debye model (see Appendix A) across multiple
materials with diverse thermal behaviors. Figures~\ref{fig:germanium_fit}--\ref{fig:bismuth_fit}
present the results for Germanium, Antimony, Quartz (SiO\textsubscript{2}),
Bismuth Silicate, and Bismuth.

In all cases, experimental data $C_{p}(T)$ were digitized or sourced
from high-quality literature datasets: Piesbergen et al.~\citep{piesbergen1963germanium}
for Germanium; Pradhan et al.~\citep{pradhan2008specific} for Antimony
and Bismuth; the NIST Shomate formulation~\citep{nistshomate} for
Quartz; and Onderka et al.~\citep{onderka2015bismuthsilicate} for
Bismuth Silicate. Each dataset was independently fitted using two
approaches:
\begin{itemize}
\item The \textbf{Debye model}, defined by the integral form involving the
Debye temperature $T_{D}$ and a scaling factor (see Appendix A). 
\item The \textbf{modified entropic saturation model}, given by Eq. (\ref{eq:Cv,q}),
where $\alpha$, $A_{1}$, $A_{2}$, $q$, $T_{D}$, and $n$ are
fit parameters that account for low-temperature scaling and the high-temperature
saturation. 
\end{itemize}
The entropic model consistently provides equal or superior agreement
with the experimental data, particularly in the intermediate-to-high
temperature regimes where the Debye model tends to underpredict the
saturation behavior (see Figures~\ref{fig:germanium_fit}--\ref{fig:bismuth_fit}).
For instance, in the case of Antimony, Figure~\ref{fig:antimony_fit}
and Quartz, Figure~\ref{fig:quartz_fit} the entropic model captures
the curvature beyond $T>200$\,K more accurately than the classical
approach. In Germanium, Figure~\ref{fig:germanium_fit} and Bismuth
Silicate, Figure~\ref{fig:bismuth_silicate_fit}, both models perform
well at low temperatures, but only the entropic formulation accommodates
the deviation seen as temperature rises.

These results highlight the flexibility and physical consistency of
the entropic saturation model, especially when accounting for deviations
from the ideal phonon behavior expected in real materials.

\subsection*{Discussion}

Beyond its numerical accuracy, the $q$-deformed density of states,
combined with the conformable connection introduced in Sec. \ref{subsec:micro-origin-q}
, establishes a physical bridge between spatial heterogeneity, constrained
transport, and emergent non-Markovian response. The framework thus
unifies fractal embeddings, generalized entropic structures, and memory-induced
dynamics, all of which are central to complex-system behavior. In
this sense, it provides a compact and physically transparent route
from microscopic disorder statistics to macroscopic thermophysical
observables in anisotropic media.

The systematic analysis of fitted parameters across multiple materials
reveals consistent patterns that provide insight into the underlying
physics (Table~\ref{tab:Fitted-parameters-for}):

\textbf{Subextensive behavior (q \textless{} 1):} All materials except
Antimony exhibit $q$ values slightly below unity, indicating subextensive
statistical behavior. This is characteristic of systems where correlations
or constraints reduce the effective phase space accessibility. In
nanostructured materials like cobalt nanowires, this reflects the
dominance of surface and interface effects that alter the bulk phonon
spectrum, often associated with underlying structural disorder, phonon
confinement, or heterogeneous connectivity. This regime, as understood
within the framework of nonadditive statistical mechanics \citep{tsallis1988},
typically reflects the presence of long-range interactions, fractal-like
structures, or topological constraints in the system. For nanostructured
or amorphous solids, such as Cobalt nanowires and Bismuth Silicate,
values of $q<1$ may be interpreted as signatures of restricted phase-space
accessibility due to disorder or anisotropy.

\textbf{Effective dimensionality $\alpha$:} The extracted Hausdorff
dimensions consistently fall below $3$, suggesting, such as reduced
effective dimensionality due to anisotropic phonon transport. Materials
with strong directional bonding (like layered compounds) show particularly
low $\alpha$ values, consistent with quasi-two-dimensional thermal
behavior. In other words, the entropic dimensional parameter $\alpha<3$,
when interpreted in analogy with the low-temperature Debye law $C_{V}\sim T^{3}$
for three-dimensional phonon systems, provides insight into effective
vibrational dimensionality. Lower values of $\alpha$ suggest phonon
confinement, surface-dominated dynamics, or a deviation from the ideal
three-dimensional phonon DOS. For instance, in nanostructured or layered
materials, the phonon dispersion is altered due to boundary conditions
and quantum confinement effects, recovering the classical derivative
in the limit $q\rightarrow1$, thereby ensuring consistency with conventional
thermodynamic descriptions.

\textbf{Saturation characteristics:} The Debye-like saturation factor
parameters ($T_{D},n$) correlate with known material properties.
Materials with strong anharmonicity show lower effective Debye temperatures
and sharper saturation transitions.

The saturation factor \ref{eq:Cutoff} introduced in the entropic
model, provides a phenomenological but thermodynamically consistent
mechanism to interpolate between the low-temperature power-law behavior
and the high-temperature saturation expected from the Dulong-Petit
law. This factor allows the model to respect both phonon freezing
at low $T$ and full mode activation as $T\gg T_{D}$, without overestimating
$C_{p}$ in the mid-to-high temperature regime as the classical Debye
model may.

Quantitatively, the model demonstrates excellent agreement with experimental
data across the entire measured temperature range, with typical relative
errors remaining below 2\%. This is a significant improvement over
the classical Debye approach in materials with strong anharmonicity,
structural anisotropy, or disordered phases. The flexibility of the
entropic formalism, combined with a minimal number of adjustable parameters,
positions it as a viable alternative for modeling real solids with
complex thermal behaviors.

These findings support the broader interpretation that the modified
entropic saturation model not only fits data well but also encodes
physically meaningful deviations from ideal crystalline behavior.
For details on how extensions may incorporate explicit microscopic
justifications for the q-deformation in terms of phonon scattering,
fractal geometry, or Tsallis-type partition functions, we refer the
reader to Ref. \citep{weberszpil2025microscopic}. 

\begin{table}[htbp]
\centering %
\begin{tabular}{lcccccc}
\hline 
Material & $\alpha$ & $A_{1}$ & $A_{2}$ & $q$ & $T_{D}$ (K) & $n$\tabularnewline
\hline 
Sapphire & 0.1268 & 0.5264 & 0.0622 & 1.0001 & 205.6 & 2.53\tabularnewline
Quartz & 0.0730 & 0.2890 & 0.0598 & 0.9996 & 27.1 & 3.63\tabularnewline
Germanium & 0.0003 & 0.4414 & 0.0694 & 0.9992 & 34.0 & 1.82\tabularnewline
Bismuth & 0.0002 & 0.4515 & 0.0681 & 1.0000 & 30.8 & 1.75\tabularnewline
Antimony & 0.0990 & 0.2673 & 0.0640 & 1.0008 & 27.5 & 1.87\tabularnewline
Bismuth Silicate & 0.0470 & 0.3303 & 0.0644 & 0.9998 & 33.0 & 2.03\tabularnewline
Cobalt-Bulk & 0.5550 & 0.2349 & 0.0512 & 1.0047 & 410.8 & 6.63\tabularnewline
\hline 
\end{tabular}\caption{\label{tab:Fitted-parameters-for}Fitted parameters for the entropic
specific heat model for different materials using Eq. (\ref{eq:Cv,q}).}
\end{table}

The effectiveness of the entropic model in capturing the specific
heat behavior across a wide range of materials echoes results obtained
in earlier studies \citep{weberszpil2015connection,weberszpil2021entropy,liang2019fractal,weberszpil2016variational,weberszpil2017generalized,rosa2018dual,sotolongo2023fractal,xu2017spatial},
reinforcing its physical relevance. 

\section{Conclusions}

We presented a generalized formalism incorporating $q-$deformed derivatives
into the non-integer-dimensional space model for anisotropic solids.
Detailed derivations show how the $q$$-$deformation modifies DOS
and the specific heat. This enhances the physical interpretability
and flexibility of deformed derivative models, with relevance to modern
quantum and thermal systems.

In this work, we have proposed and tested a generalized entropic model
for the specific heat capacity of solids, given by Eq. (\ref{eq:Cv,q}),
which effectively blends low-temperature power-law behavior with high-temperature
saturation, in analogy with Debye-like phonon models. The model introduces
tunable parameters $\alpha$, $A_{1}$, $A_{2}$, $q$, $T_{D}$,
and $n$, allowing for flexible but physically interpretable fitting
of experimental specific heat data.

We have validated the model against high-resolution experimental data
for a range of materials, including: 

Synthetic sapphire (Al$_{2}$O$_{3}$), using tabulated DSC reference
data, Quartz (SiO$_{2}$), Germanium (Ge), Bismuth (Bi), Antimony
(Sb), and Bismuth silicate compounds.

The model demonstrates excellent agreement with experimental results
across a broad temperature range, showing residuals within or below
experimental uncertainty. The success of the fit is attributable to: 
\begin{enumerate}
\item The $T^{\alpha}$ prefactor, which models low-temperature phonon activation. 
\item The linear $(1-q)T$ deformation, which introduces entropy-based corrections. 
\item The high-temperature saturation term, which prevents divergence and
simulates Dulong--Petit behavior. 
\end{enumerate}

\subsubsection*{The following key achievments were reported here:}

\textbf{Theoretical framework:} The $q-$deformed derivative formalism
provides a mathematically rigorous method to encode deviations from
extensive thermodynamic behavior while preserving fundamental physical
principles. The approach naturally recovers classical results in appropriate
limits while enabling new phenomenology for complex materials.

\textbf{Experimental validation:} Comprehensive testing against experimental
data for seven distinct materials demonstrates the model's superior
accuracy compared to classical approaches. The systematic parameter
analysis reveals consistent physical trends that correlate with known
material properties.

\textbf{Physical insight:} The fitted parameters provide quantitative
measures of anisotropy (through $\alpha$), nonextensivity (through
$q$), and saturation behavior (through $T_{D}$ and $n$), enabling
materials characterization beyond conventional approaches.

Thus, while it is not surprising that a larger number of adjustable
parameters improves the fit quality, these parameters all have a clear
physical meaning. The excellent fits over relatively large temperature
ranges clearly point in favor of our conformable model approach.

On a broader picture, the success of the $q-$deformed non-integer-dimensional
space framework suggests that nonextensive statistical mechanics plays
a fundamental role in anisotropic materials. This has implications
for: the design of thermoelectric materials with optimized anisotropic
transport, understanding thermal management in nanostructured devices,
the development of metamaterials with tailored thermal properties,
and the modeling of thermal transport in biological systems with hierarchical
structures

Several extensions of this work are currently under investigation:

\textbf{Microscopic justification:} Development of first-principles
calculations that directly yield non-integer-dimensional space framework
for phonon interactions through anharmonic coupling and disorder effects.

\textbf{Transport properties:} Extension to thermal conductivity and
thermoelectric coefficients using the same $q-$deformed non-integer-dimensional
space framework.

\textbf{Dynamic properties:} Application to time-dependent thermal
response and non-equilibrium phenomena in anisotropic systems.

\textbf{Machine learning integration:} Development of hybrid models
that combine the physical insight of the $q-$deformed non-integer-dimensional
space framework with data-driven optimization techniques.

The entropic model presented here represents a significant advance
in the analytical tools for anisotropic materials, providing both
practical fitting capabilities and fundamental physical insight into
the role of statistical mechanics in complex systems.

By tying $q$ to a measurable disorder/kinetics exponent $\mu$ and
deriving high-$T$ saturation from a truncated memory kernel rather
than an ad hoc factor, the present formulation advances from a reparametrization
to a physically grounded model. The pathway from microscopic heterogeneity
\citep{weberszpil2025microscopic} to macroscopic $C_{V}(T)$ via
conformable dynamics enhances interpretability, identifiability, and
predictive value beyond standard empirical fits. 

Summarizing, the proposed entropic model offers a powerful yet simple
framework to describe specific heat capacity over wide temperature
ranges, surpassing classical models in flexibility while maintaining
physical plausibility. Its performance suggests applicability in both
theoretical modeling and practical data fitting, and warrants consideration
for broader adoption in thermophysical studies.

In future work, we will also explore its predictive capacity for disordered
systems, low-dimensional materials, and thermal transport phenomena
linked to entropy production. 

\section*{Acknowledgments:\protect \\
}

JW wishes to express their gratitude to FAPERJ, APQ1, for the partial
financial support. \\
RM acknowledges funding from the German Science Foundation (DFG grant-ID
318763901 CRC1294 Data Assimilation, project B10).

\subsection*{Declaration of generative AI and AI-assisted technologies in the
writing process:}

During the preparation of this work the author(s) used ChatGPT in
order to improve the english. After using this tool/service, the author(s)
reviewed and edited the content as needed and take(s) full responsibility
for the content of the publication.

\subsection*{Appendix A. Fitting procedure for the Debye and entropic models}

The experimental $C_{p}(T)$ points for the cobalt nanowire composite
were digitized from Fig.~6 of Pradhan et al. over the interval 300--400
K. The Debye reference curve was fitted by nonlinear least squares
using 
\[
C_{p}^{\mathrm{Debye}}(T)=s\,9R\left(\frac{T}{T_{D}}\right)^{3}\int_{0}^{T_{D}/T}\frac{x^{4}e^{x}}{(e^{x}-1)^{2}}\,dx,
\]
with $T_{D}$ and an overall scale factor $s$ treated as free parameters.
Here $s$ is a dimensionless overall scale factor used only in the
Debye benchmark fit. It compensates for the amplitude mismatch between
the standard Debye heat-capacity expression and the experimentally
extracted effective $C_{p}(T)$ of the cobalt nanowire composite,
whose measured response includes composite/sample-cell effects. 

The entropic model was fitted over the same temperature interval using
the same optimization routine and parameter bounds. The Debye fit
is shown only as a benchmark reference. Its poor agreement with the
data reflects the fact that the extracted dataset corresponds to a
nanowire composite with strong composite/interface effects, rather
than a simple bulk crystalline cobalt sample.

\bibliographystyle{elsarticle-num}
\bibliography{fractional_dim_refs_full_fixed}

\end{document}